\documentclass{ifacconf}

\usepackage{graphicx}      % include this line if your document contains figures
\usepackage{natbib}        % required for bibliography
\usepackage[utf8]{inputenc}   % This two packages are used for \v and \'
\usepackage[T1]{fontenc}      % same as previous

\usepackage{subfigure, amsmath, rotating}
\usepackage{array, multirow}
\usepackage{amsbsy, amsfonts, graphics}
\usepackage{algorithmic}
\usepackage{algorithm}
\usepackage{epstopdf}
\usepackage{mathrsfs}
\usepackage{caption}
\usepackage{amssymb}
\usepackage{amsmath,bm}
\usepackage{subfiles}
\usepackage{mathtools}

\usepackage{tikz}
\usepackage{float}
\usetikzlibrary{calc,patterns,decorations.pathmorphing,decorations.markings}
\usepackage{soul}
\usepackage{cancel}
\usepackage{todonotes}

\usepackage{subfig}
\usepackage[normalem]{ulem} % 给删除线准备的，haing [normalem] makes \emph{} to be italic not underline.
\usepackage{units}
\newcommand{\bea}{\begin{eqnarray}}
\newcommand{\eea}{\end{eqnarray}}

\definecolor{mygreen}{RGB}{0,180,0}

\usepackage{wrapfig}

\makeatletter
\renewcommand{\maketag@@@}[1]{\hbox{\m@th\normalsize\normalfont#1}}%
\makeatother

\newtheorem{ass}{Assumption}
\newtheorem{con}{Condition}
\newtheorem{theorem}{Theorem}
\newtheorem{remark}{Remark}

\begin{document}
\begin{frontmatter}

\title{Observer Design for Networked Linear Systems with Fast and Slow Dynamics under Measurement Noise\thanksref{footnoteinfo}} 

\thanks[footnoteinfo]{
This work was supported by the Australian Research Council under the Discovery Project DP200101303, the France Australia collaboration project IRP-ARS CNRS and the ANR COMMITS
ANR-23-CE25-0005, and by ANID through the FONDECYT Initiation grant 11250715. E-mail corresponding author: weixuanw@student.unimelb.edu.au.
}

\author[Melbourne]{W. Wang}%\ead{weixuanw@student.unimelb.edu.au},
\author[Chile]{A. I. Maass}%\ead{alejandro.maass@uc.cl},
\author[Melbourne]{D. Ne\v{s}i\'{c}}%\ead{dnesic@unimelb.edu.au},
\author[Melbourne]{Y. Tan}%\ead{yingt@unimelb.edu.au},
\author[France]{R. Postoyan}%\ead{romain.postoyan@univ-lorraine.fr},
\newline
\author[Netherlands]{W.P.M.H. Heemels}%\ead{w.p.m.h.heemels@tue.nl}

\address[Melbourne]{Faculty of Engineering and Information Technology, The University of Melbourne, Parkville, 3010, Victoria, Australia}
\address[Chile]{Department of Electrical Engineering, Pontificia Universidad Cat\'olica de Chile, Santiago, 7820436, Chile}
\address[France]{Universit\'e de Lorraine, CNRS, CRAN, F-54000 Nancy, France}
\address[Netherlands]{Department of Mechanical Engineering, Eindhoven University of Technology, The Netherlands}

\begin{abstract}                % Abstract of 50--100 words
This paper addresses the emulation-based observer design for networked control systems (NCS) with linear plants that operate at two time scales in the presence of measurement noise. The system is formulated as a hybrid singularly perturbed dynamical system, enabling the systematic use of singular perturbation techniques to derive explicit bounds on the maximum allowable transmission intervals (MATI) for both fast and slow communication channels. Under the resulting conditions, the proposed observer guarantees that the estimation error satisfies a global exponential derivative-input-to-state stability (DISS)-like property, where the ultimate bound scales proportionally with the magnitudes of the measurement noise and the time derivative of the control input. The effectiveness of the approach is illustrated through a numerical example.
\end{abstract}

\begin{keyword}
Networked control systems, observer design, singular perturbed systems, hybrid dynamical systems
\end{keyword}

\end{frontmatter}
%===============================================================================

\section{Introduction}
NCS, namely control systems whose loops are closed through packet-based communication networks, are popular across many domains. In some of these applications, the controlled systems feature multi–time-scale dynamics. For such systems, intuitively , signals that vary slowly can be sampled and transmitted at lower rates than those that change rapidly. However, most foundational works on NCS, e.g., \cite{carnevale_stability,yuksel2013stochastic}, ignore the presence of multiple time scales and instead assume a uniform transmission rate, which lead unnecessarily high transmission rates for slow-varying signals and may impose avoidable communication burdens.
Recent studies e.g., \cite{Romain_ETC,SPNCS,Single_channel_NCS_CDC,Weixuan_Journal}, address the control of NCS with multiple time scales by applying singular perturbation techniques \citep{nonlinear_systems_Khalil} to mitigate redundant transmissions of slow-varying signals and handle the inherent stiffness problem, leading to so called singularly perturbed NCS (SPNCS).

In addition to the challenges introduced by multiple time scales, many practical systems also involve internal states that cannot be directly measured. For example, in battery management, the state-of-charge and state-of-health are not directly measured and must instead be estimated. In such scenarios, observers are indispensable tools for estimating the unmeasurable states. Moreover, electrochemical, thermal and aging dynamics of battery evolve on multiple time scales, which naturally lead to the use of multi-time-scale observers \citep{zou2016multi}. Extensive research has been conducted on the multi-time-scale observer design for both continuous-time and sampled-data singularly perturbed systems, e.g., \cite{lin2010composite,yang2016observer,ferrante2021observer}, while several studies also investigated observers for NCS, e.g., \cite{postoyan2011framework,wang2017observer}. Despite these advances, the development of multi–time–scale observers for SPNCS that explicitly account for measurement noise, which frequently occurs in practice, remains unexplored.

In this work, we investigate the multi-time-scale observer design for SPNCS with linear plants, where the plant input, and both fast- and slow-varying measurements, are transmitted over a packet-based network subject measurement noise. 
Despite the linearity of the plant and observer, the problem remains nontrivial due to the potentially time-varying and nonlinear scheduling protocol, the hybrid nature of the closed-loop system, and network imperfections such as non-equidistant sampling.

Following an emulation-based approach, a Luenberger observer is first designed in continuous time while enhancing robustness against network-induced errors under the adopted uniformly globally exponentially stable (UGES) scheduling protocols \citep{dragan_stability}, with the resulting conditions expressed as linear matrix inequalities (LMI). This observer is then implemented over the network and the estimates on the maximum allowable transmission intervals (MATI) of slow and fast signals are provided, respectively.
Under these conditions, the estimation error satisfies a global exponentially DISS-like property, in the sense of \cite{DISS}. In particular, ultimate bound on the estimation error is proportional to the magnitudes of measurement noise and the time derivative of the plant input. 

This work provides two main contributions. First, it introduces a systematic observer design framework for SPNCS with linear plants via LMI. In particular, the proposed observer explicitly incorporates key network-induced effects, including sampling and scheduling protocols. By doing so, it enables the derivation of observer gains that make the observer robust against network-induced errors and thereby maximize the MATI estimates for both slow- and fast-varying signals. This contribution provides a unified and computationally tractable method for observer synthesis in multi-time-scale NCS, where network constraints play a critical role. 
Second, the paper establishes a robustness analysis framework for observer design in SPNCS with linear plants subject to measurement noise and plant input. Because noisy measurements are sampled and transmitted over the network, the communication process inherently attenuates high-frequency noise. Following the idea in \cite{scheres2024robustifying}, by analysing the received noise (i.e., the noise after it passes through the network), we remove the dependence of the ultimate bound on the time derivative of the measurement noise, leading to a more practical and implementable robustness guarantee.

\newcommand{\esssup}{\mathop{\mathrm{ess.sup}}\limits}
\newcommand{\supNew}{\mathop{\mathrm{sup}}\limits}
\textbf{Notation:} 
% Let $\mathbb{R}_{\geq 0} \coloneqq [0,\infty)$ and for any integer $n$, $\mathbb{Z}_{\geq n} \coloneqq \{n, n+1, n+2, \cdots \}$.
% For vectors $v_i\in \mathbb{R}^n$, $i\in \{1,2,\cdots, N\}$, we denote the vector $[v_1^\top \; v_2^\top \; \cdots \; v_N^\top]^\top$ by $(v_1, v_2, \cdots, v_N)$, and inner product by $\left< \cdot , \cdot \right>$. 
Let $\mathbb{R}_{\geq 0} \coloneqq [0,\infty)$ and for any integer $n$, $\mathbb{Z}_{\geq n} \coloneqq \{n, n+1, \cdots \}$.
For vectors $v_i\in \mathbb{R}^n$, $i\in \{1,2,\cdots, N\}$, we denote the vector $[v_1^\top \; v_2^\top \; \cdots \; v_N^\top]^\top$ by $(v_1, v_2, \cdots, v_N)$, and inner product by $\left< \cdot , \cdot \right>$. 
Given a vector $x\in \mathbb{R}^{n_x}$ and a non-empty closed set $\mathcal{A} \subseteq \mathbb{R}^{n_x}$, the distance from $x$ to $\mathcal{A}$ is denoted by $|x|_\mathcal{A} \coloneqq \min_{y\in \mathcal{A}}|x-y|$. 
Given a hybrid arc $u: \text{dom}(u) \rightarrow \mathcal{U}$ \cite{gosate12}, define 
 $
     \|u\|_{(t_1,j_1)} \coloneqq \max \Big\{
  \esssup_{(t,j)\in \mathrm{dom}(u)\, \setminus \,\Gamma_u(u), \,(t,j) \preceq (t_1,j_1)}
  |u(t,j)|, $ $
  \supNew_{(t,j)\in \Gamma_u(u), \ (t,j) \preceq (t_1,j_1)}
  |u(t,j)| \Big\}
$
where $\Gamma_u(u)$ denotes the set of all $(t,j)\in \text{dom}(u)$ such that $(t, j+1) \in \text{dom}(u)$.
For a real symmetric matrix $P$, we denote its maximum and minimum eigenvalues by $\lambda_{\text{max}}(P)$ and $\lambda_{\text{min}}(P)$ respectively. The logic AND operator is denoted by $\wedge$.

\section{Problem setting}
\label{Section Problem setting - Observer}
\begin{figure}[ht]
    \centering
    \includegraphics[width = 0.75\linewidth]{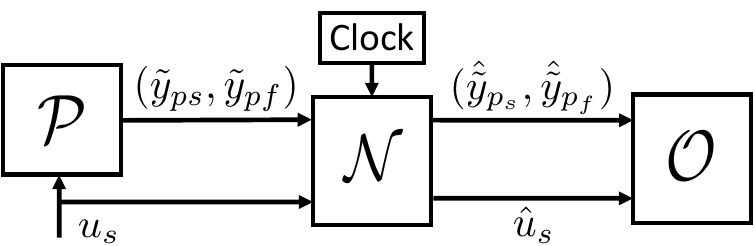}
    \caption{Block diagram of observer for SPNCS}
    \label{fig: Block Diagram - Observer}
\end{figure}

This section outlines the structure of the plant $(\mathcal{P})$, observer $(\mathcal{O})$ and the communication network $(\mathcal{N})$ as depicted in Figure \ref{fig: Block Diagram - Observer}.

\subsection{Plant and observer}
Consider the following linear singularly perturbed plant
\begin{equation}
    \mathcal{P}:
    \begin{cases}
    \begin{aligned}
    \dot x_p &= A_{11}x_p + A_{12}z_p + B_1 u_s\\
    \epsilon \dot z_p &=  A_{21}x_p + A_{22}z_p + B_2 u_s\\
    \tilde{y}_{p_s} &= C_{1s}x_p + v_1, \\
    \tilde{y}_{p_f} &= C_{2s}x_p + C_{2f}z_p + v_2,
    \end{aligned}
    \end{cases} 
    \label{eqn: plant}
\end{equation}
where $0 < \epsilon \ll 1$ is a known small parameter, $A_{22}$ is invertible and the pairs $(A_{11}, C_{1s})$ and $(A_{22}, C_{2f})$ are observable. Here, $x_p \in \mathbb{R}^{n_x}$ and $z_p\in \mathbb{R}^{n_z}$ denote the slow and fast states of the plant, respectively; $\tilde y_{p_s}\in \mathbb{R}^{n_{y_s}}$ and $\tilde y_{p_f} \in \mathbb{R}^{n_{y_f}}$ represent the slow and fast measured outputs; and $v_1, v_2\in \mathcal{L}_\infty$ are measurement noise, where $n_{x_p}, n_{z_p}, n_{y_s}, n_{y_f} \in \mathbb{Z}_{\geq 1}$.
Moreover, $u_s$ is the ``slow'' input, and we assume it is differentiable with bounded derivative (in an $\mathcal{L}_\infty$ sense) when $\epsilon$ approaches zero (i.e., $u_s$ is absolutely continuous). This assumption is well justified in engineering practice, as most actuators cannot change instantaneously, i.e., they are rate-limited due to physical constraints such as inertia and friction, electrical restrictions on current and voltage, and safety considerations that prevent excessive accelerations.

For given $\epsilon$ in \eqref{eqn: plant}, the following observer is introduced:
\begin{equation}
    \mathcal{O}:
    \begin{cases}
    \begin{aligned}
    \dot x_o =& A_{11}x_o + A_{12}z_o + B_1\hat{u}_s \\
                &+ L_{1s}(\hat{\tilde y}_{p_s}-\hat{y}_{o_s}) + L_{1f}(\hat{\tilde y}_{p_f} - \hat{y}_{o_f})\\
    \epsilon \dot z_o =&  A_{21}x_o + A_{22}z_o + B_2\hat{u}_s\\
                        &+ L_{2s}(\hat{\tilde y}_{p_s}-\hat{y}_{o_s}) + L_{2f}(\hat{\tilde y}_{p_f} - \hat{y}_{o_f})\\
    y_{o_s} =& C_{1s}x_o, \\
    y_{o_f} =& C_{2s}x_o + C_{2f}z_o,
    \end{aligned}
    \end{cases}
    \label{eqn: observer}
\end{equation}
where $x_o$ and $z_o $ are the estimated states. The variables $\hat{\tilde  y}_{p_s}$ and $\hat{\tilde y}_{p_f}$ denote the most recently received noisy plant outputs transmitted over the network, while $\hat{y}_{o_s}$ and $\hat{y}_{o_f}$ are introduced as network-induced counterparts to ${y}_{o_s}$ and ${y}_{o_f}$, respectively, following the approach suggested in \cite{wang2017observer}. 
For analysis purposes, we introduce  auxiliary noise-free outputs $y_{p_s} = C_{1s}x_p$ and $y_{p_f} = C_{2s}x_p + C_{2f}z_p$.
The observer gains $L_{1s}$, $L_{1f}$, $L_{2s}$ and $L_{2f}$ will be designed such that estimation error dynamics are robust to the network-induced error.
%are designed to ensure exponential stability of the estimation error under perfect communication and measurement, i.e., when $\hat{\tilde  y}_{p_s} = y_{p_s}$, $\hat{\tilde y}_{p_f} = \hat{\tilde y}_{p_f}$, $\hat{y}_{o_s} = y_{o_s}$ and $\hat{y}_{o_f} = y_{o_f}$.
%,postoyan2014tracking}.
In this work, we assume that zero-order-hold is applied to $\hat{\tilde y}_{p_s}$, $\hat{\tilde y}_{p_f}$ and $\hat{u}_s$ (i.e., $\dot{\hat{\tilde y}}_{p_s} =  \dot{\hat{\tilde y}}_{p_f}= \dot{\hat{u}}_s = 0$) between successive transmission instances.

\subsection{Network}
We define the \emph{measured} network-induced errors as %$\tilde e_{p_s} \coloneqq \hat{\tilde y}_{p_s} - \tilde y_{p_s}$, $\tilde e_{p_f} \coloneqq \hat{\tilde y}_{p_f} - \tilde y_{p_f}$ and $e_{u_s} \coloneqq \hat{u}_s - u_s$.
\begin{equation}
    \tilde e_{p_s} \coloneqq \hat{\tilde y}_{p_s} - \tilde y_{p_s},\quad
    \tilde e_{p_f} \coloneqq \hat{\tilde y}_{p_f} - \tilde y_{p_f},\quad
    e_{u_s} \coloneqq \hat{u}_s - u_s.
    \label{eqn: real plant network-induced errors}
\end{equation}
As shown in Figure \ref{fig: Block Diagram - Observer}, we consider the scenario in which $\tilde y_{p_s}$, $\tilde y_{p_f}$ share a common communication channel, while $u_s$ is transmitted over a dedicated channel. Each channel may consist of several \emph{network nodes}, with each node exclusively handling either slow or fast signals. 
This configuration guarantees that slow and fast signals are never transmitted simultaneously. 
In particular, suppose there are $\ell$ slow nodes in the channel of plant output, then we can decompose $\tilde e_{p_s} = (\tilde e_{p_s}^1, \cdots, \tilde e_{p_s}^{\ell})$ with each $k^{\text{th}}$ component  having dimension $n_{k}$ corresponds to the $k^{\text{th}}$ slow node, while satisfying $\sum_{k=1}^{\ell} n_k = n_{y_s}$.

Let $\mathcal{T} \coloneqq \{t_1, t_2, \cdots \}$ represent the set of all transmission instants. Define the subset $\mathcal{T}^s \coloneqq \{t_1^s, t_2^s, \cdots \}$ as the instants when a slow node accesses the network, and let $\mathcal{T}^f \coloneqq \mathcal{T} \setminus \mathcal{T}^s = \{t_1^f, t_2^f, \cdots \}$ be the instants when a fast node transmits. For any $k \in \mathbb{Z}_{\geq 1}$, the transmission intervals satisfy
\begin{subequations}
    \begin{align}
    &\tau_{\text{miati}}^s \leq t_{k+1}^s - t_k^s \leq \tau_{\text{mati}}^s, && \forall t_k^s,t_{k+1}^s\in \mathcal{T}^s,  \label{eqn: timer eqn1 - Observer}
    \\
    &\tau_{\text{miati}}^f \leq t_{k+1}^f - t_k^f \leq \tau_{\text{mati}}^f ,  &&\forall t_k^f, t_{k+1}^f  \in \mathcal{T}^f,  
    \label{eqn: timer eqn2 - Observer}
    \\
    &\tau_{\text{miati}}^f \leq t_{k+1} - t_k,  && \forall t_k, t_{k + 1} \in \mathcal{T}. \label{eqn: timer eqn3 - Observer}
    \end{align}
    \label{eqn: Stefan timer - Observer}%
\end{subequations}
\noindent Here, $0<\tau_{\text{miati}}^f\leq \tau_{\text{mati}}^f$ denote the minimum allowable transmission interval (MIATI) and MATI for consecutive fast transmissions, respectively, and $\tau_{\text{miati}}^s$ and $\tau_{\text{mati}}^s$ are defined similarly for slow transmissions. Since a slow transmission of $\tilde y_{p_s}$ may occur between two consecutive fast transmissions of $\tilde y_{p_f}$, the condition $\tau_{\text{miati}}^f \leq  \tfrac{1}{2}\tau_{\text{mati}}^f$
% \begin{equation*}
%     \tau_{\text{miati}}^f \leq  \tfrac{1}{2}\tau_{\text{mati}}^f
%     %\label{eqn: condition on miati^f}
% \end{equation*}
is necessary for \eqref{eqn: timer eqn2 - Observer} and \eqref{eqn: timer eqn3 - Observer} to be valid.

% We define the \emph{measured} network-induced errors as %$\tilde e_{p_s} \coloneqq \hat{\tilde y}_{p_s} - \tilde y_{p_s}$, $\tilde e_{p_f} \coloneqq \hat{\tilde y}_{p_f} - \tilde y_{p_f}$ and $e_{u_s} \coloneqq \hat{u}_s - u_s$.
% %
% \begin{equation*}
%     \tilde e_{p_s} \coloneqq \hat{\tilde y}_{p_s} - \tilde y_{p_s},\quad
%     \tilde e_{p_f} \coloneqq \hat{\tilde y}_{p_f} - \tilde y_{p_f},\quad
%     e_{u_s} \coloneqq \hat{u}_s - u_s.
%\end{equation*}
Since the noisy measurements are sampled and transmitted, this process naturally filters out high-frequency noise. As a result, we only need to address the transmitted noise, rather than the original $v_1$ and $v_2$.
%\cyan{Since the noisy measurements are sampled and transmitted, the effective noise is not $v_1$ and $v_2$ but their sampled and transmitted versions.} 
For analytical purposes, we introduce the \emph{unmeasured} noise-free variables $\hat{y}_{p_s}$, $\hat{y}_{p_f}$, $\hat v_1$ and $\hat v_2$, where $\hat{\star}$ denotes the most recently received $\star$ for all $\star \in \{y_{p_s}, y_{p_f}, v_1, v_2 \}$ like in \cite{scheres2024robustifying}. Thus, we have $\hat{\tilde y}_{p_s} = \hat{y}_{p_s} + \hat v_1$ and $\hat{\tilde y}_{p_f} = \hat{y}_{p_f} + \hat v_2$.
 We define the \emph{ideal} network-induced error %$e_{p_s} \coloneqq \hat{y}_{p_s} - y_{p_s}$ and $e_{p_f} \coloneqq \hat{y}_{p_f} - y_{p_f}$, 
\begin{equation*}
    e_{p_s} \coloneqq \hat{y}_{p_s} - y_{p_s}, \quad e_{p_f} \coloneqq \hat{y}_{p_f} - y_{p_f},
\end{equation*}
which leads to %$\tilde e_{p_s} = e_{p_s} + \hat v_1 - v_1$ and $\tilde e_{p_f} = e_{p_f} + \hat v_2 - v_2.$
\begin{align*}
   \tilde e_{p_s} = e_{p_s} + \hat v_1 - v_1, \quad \tilde e_{p_f} = e_{p_f} + \hat v_2 - v_2. 
\end{align*}

By adopting a similar approach introduced in \cite{postoyan2014tracking}, we introduce artificial network-induced errors on the observer outputs, defined as %$e_{o_s} \coloneqq \hat{y}_{o_s} - y_{o_s}$ and $e_{o_f} \coloneqq \hat{y}_{o_f} - y_{o_f}$. 
\begin{equation*}
    e_{o_s} \coloneqq \hat{y}_{o_s} - y_{o_s}, \quad e_{o_f} \coloneqq \hat{y}_{o_f} - y_{o_f}.
\end{equation*}
The variables $\hat{y}_{o_s}$ and $\hat{y}_{o_f}$ are designed such that $e_{o_s}$ and $e_{o_f}$ evolve along the same vector field as $\tilde e_{p_s}$ and $\tilde e_{p_f}$ at transmission instants, respectively. 
To be explicit, we decompose the error and noise vectors as $e_{p_s} = (e_{p_s}^1, \cdots, e_{p_s}^{\ell})$, $e_{o_s} = (e_{o_s}^1, \cdots, e_{o_s}^{\ell})$ and $\hat{v}_1 = (\hat v_1^1, \cdots, \hat v_1^{\ell})$ with each $k^{\text{th}}$ component having dimension $n_{k}$ (i.e., same as $\tilde e_{p_s}$). When the slow node $j \in \{1, \cdots, \ell \}$ in $\tilde e_{p_s}$ accesses the channel at time $t_k^s \in \mathcal{T}^s$, the associated vectors are reset according to
\begin{equation}
    (\tilde e_{p_s}^{k^+}, e_{p_s}^{k^+}, e_{o_s}^{k^+}, \hat v_1^{k^+}) = \left\{
    \begin{aligned}
        &(0,0,0,v_1^k) &&\text{if } k = j ,\\
        & ( \tilde e_{p_s}^{k}, e_{p_s}^{k}, e_{o_s}^{k}, \hat v_1^{k}) &&\text{if } k \neq j,
    \end{aligned}
    \right. \label{eqn: Update rules - Observer}
\end{equation}
for all $k\in \{1, \cdots, \ell \}$. Furthermore, we define the network-induced error on the \emph{ideal} output estimation error as %$e_{y_s} \coloneqq e_{p_s} - e_{o_s} $ and $e_f \coloneqq e_{p_f} - e_{o_f}$ 
\begin{equation}
    e_{y_s} \coloneqq e_{p_s} - e_{o_s}, \quad e_f \coloneqq e_{p_f} - e_{o_f},
    \label{eqn: ideal errors}
\end{equation}
so that $\hat{\tilde y}_{p_s}-\hat{y}_{o_s} = y_{p_s}-y_{o_s} + e_{y_s} + \hat v_1$ and $\hat{\tilde y}_{p_f}-\hat{y}_{o_f} = y_{p_f}-y_{o_f} + e_{f} + \hat v_2$. 
%

%Before we present the behaviour of the system at transmission times, we introduce some useful notation regarding the variables: $x\coloneqq (x_p,x_c)\in\mathbb{R}^{n_x}$, $z \coloneqq ( z_p, z_c) \in \mathbb{R}^{n_z}$, $e_s \coloneqq ( e_{y_s} , e_{u_s})\in \mathbb{R}^{n_{e_s}}$ and $e_f \coloneqq (e_{y_f} , e_{u_f}) \in \mathbb{R}^{n_{e_f}}$, with $n_x\coloneqq n_{x_p}+n_{x_c}$,  $n_z\coloneqq n_{z_p}+n_{z_c}$, $n_{e_s}\coloneqq n_{y_s}+n_{u_s}$ and  $n_{e_f}\coloneqq n_{y_f}+n_{u_f}$. 
 
At each transmission instance $t_k^s \in \mathcal{T}^s$ of a slow node, the network-induced errors are updated as follows:
$\tilde e_{p_s}^+ = h_{\tilde{p}s}(k, \tilde e_{p_s})$, 
$e_{us}^+ = h_{us}(k, e_{us})$,
$e_{y_s}^+ = h_{y_s}(k, e_{y_s},e_{p_s})$,
$\tilde e_{p_f}^+ = \tilde e_{p_f}$, 
$ e_{o_f}^+ = e_{o_f}$,
$e_f^+ = e_{f}$.
In addition, the variable $\hat v_1$ is updated via $\hat v_1^+ = h_{v1}(k, \hat v_1, v_1, \tilde e_{p_s})$.
Here, $h_{\tilde{p}s}$ and $h_{us}$ model the scheduling protocol that manage the slow updates of $\tilde y_{p_s}$ and $u_s$ as in \cite{dragan_stability}, respectively, while $h_{y_s}$ and $h_{v1}$ are defined based on the update rule \eqref{eqn: Update rules - Observer} and definition of $e_{y_s}$.
Similarly, for each transmission time $t_k^f \in \mathcal{T}^f$ of a fast node, the updates are given by: 
$\tilde e_{p_s}^+ = \tilde e_{p_s}$,
$e_{us}^+ = e_{us}$,
$e_{y_s}^+ = e_{y_s}$,
$\tilde e_{p_f}^+ = h_{\tilde{p}f}(\kappa_f,\tilde e_{p_f})$,
$e_{f}^+ = h_{f}(\kappa_f, e_{f},\tilde e_{p_f})$,
and $\hat v_2^+ = h_{v_2}(\kappa_f, \hat v_2, v_2,\tilde e_{f})$.

\section{A hybrid model for the SPNCS}
\label{Section Modeling - Observer}
\begin{figure*}[!htp]
	\hrule
    \footnotesize % You can change this to \small \footnotesize, \scriptsize, or \tiny
    \begin{equation}
        \begin{aligned}
                &\begin{aligned}
                    f_{\delta_x}(\delta_x,e_{y_s}, e_{u_s},\delta_y, e_f,\hat{v}) \coloneqq&  
                (A_{11}-L_{1s}C_{1s} - L_{1f}C_{2s}) \delta_x + (A_{12} - L_{1f}C_{2f})(\delta_y + \overline{H}(\delta_x, e_{y_s}, e_{u_s},\hat{v}))
                \\
                &+ B_1 e_{u_s} + L_{1s} (e_{y_s}+\hat v_1) + L_{1f}(e_f+\hat v_2)
                \end{aligned}
            \\
            &f_{e_{y_s}}(\delta_x,e_{y_s}, e_{u_s},\delta_y, e_f,\hat{v}) \coloneqq 
            C_{1s}f_{\delta_x}(\delta_x,e_{y_s}, e_{u_s},\delta_y, e_f,\hat{v})
            \\
            &F_{\xi_s}(\xi_s,\delta_y,e_f,\hat{v},\dot{u}_s) \coloneqq \big( f_{\delta_x}(\delta_x,e_{y_s}, e_{u_s},\delta_y, e_f,\hat{v}),  f_{e_{y_s}}(\delta_x,e_{y_s}, e_{u_s},\delta_y, e_f,\hat{v}), -\dot{u}_s, 1, 0 ,0 \big)
            \\
            &g_{\delta_y}(\delta_x,e_{y_s}, e_{u_s},\delta_y, e_f,\hat{v}, \epsilon) \coloneqq 
            (A_{22} - L_{2f}C_{2f}) \delta_y  + L_{2f}e_f - \epsilon \tfrac{\partial \overline{H}}{\partial \xi_s} F_{\xi_s}(\xi_s,\delta_y, e_f, \hat{v}, \dot{u}_s)
            \\
            &g_{e_f}(\delta_x,e_{y_s}, e_{u_s},\delta_y, e_f,\hat{v}, \epsilon) = \epsilon C_{2s}f_{\delta_x}(\delta_x,e_{y_s}, e_{u_s},\delta_y, e_f,\hat{v}) + C_{2f}\big((A_{22} - L_{2f}C_{2f}) \delta_y +  L_{2f}e_f    \big) 
            \\
             &F_{\xi_f}(\xi_s,\xi_f,\hat{v}, \dot{u}_s, \epsilon) \coloneqq \big( g_{\delta_y}(\delta_x,e_{y_s}, e_{u_s},\delta_y, e_f,\hat{v}, \epsilon), g_{e_f}(\delta_x,e_{y_s}, e_{u_s},\delta_y, e_f,\hat{v}, \epsilon), 1, 0, 0 \big)
        \\
            &G_{s,1}(\xi_s, \zeta_s,v_1) \coloneqq \big(\delta_x,h_{ys}(\kappa_s, e_{y_s}, \tilde e_{p_s}),h_{us}(\kappa_s, e_{u_s}),0 , \kappa_s + 1, h_{v_1}(\kappa_s, \hat v_1, v_1, \tilde e_{p_s}) ,x_p , h_{\tilde{p}s}(\kappa_s, \tilde e_{p_s})\big) 
            \\
            &G_{s,2}(\xi,v_1) \coloneqq \big(h_y(\kappa_s, e_{y_s}, e_{u_s}, \tilde e_{p_s},\delta_y, \hat v_1, v_1), e_f,\tau_f,  \kappa_f, \hat v_2 , z_p, \tilde e_{p_f}  \big)
            % \\
            % G_s(\xi,v_1) &\coloneqq \big(\delta_x,h_{ys}(\kappa_s, e_{y_s}, \tilde e_{p_s}),h_{us}(\kappa_s, e_{u_s}),0 , \kappa_s + 1, h_{v_1}(\kappa_s, \hat v_1, v_1, \tilde e_{p_s}) ,x_p, h_{\tilde{p}s}(\kappa_s, \tilde e_{p_s}), \red{h_y(\kappa_s, e_{y_s}, e_{u_s}, \tilde e_{p_s},\delta_y, \hat v_1, v_1)}, e_f,\tau_f,  \kappa_f, \hat v_2 , z_p, \tilde e_{p_f}  \big)
            \\
            &\begin{aligned}
                G_f(\xi, v_2) \coloneqq \big(&\delta_x,e_{y_s}, e_{u_s}, \tau_s, \kappa_s,\hat v_1, x_p, \tilde e_{p_s},  
                \delta_y, h_{f}(\kappa_f, e_f, \tilde e_{p_f}), 0, \kappa_f + 1 ,  h_{v_2}(\kappa_f, \hat v_2, v_2, \tilde e_f),z_p, h_{\tilde{p}f}(\kappa_f, \tilde e_{p_f}) \big)
            \end{aligned}
    \end{aligned} \label{eqn: Functions}
    \end{equation}
    \footnotesize
	\hrule
    %\vspace{-15pt}
\end{figure*}
We extend the standard singular perturbation technique \cite[Section 11.5]{nonlinear_systems_Khalil} to hybrid systems. Specifically, we begin by determining the quasi-steady state followed by a coordinate transformation. We then investigate the stability of the SPNCS by analyzing both the \emph{boundary-layer} (fast) and the \emph{reduced} (slow) system.

\subsection{Change of coordinates}
Define the estimation errors on the slow and fast states as $\delta_x \coloneqq x_o - x_p$ and $\delta_z \coloneqq z_o - z_p$, respectively. Let 
\begin{equation}
    v \coloneqq ( v_1, v_2),\quad \hat{v} \coloneqq (\hat v_1, \hat v_2).
    \label{eqn: definition of v and v_hat}
\end{equation}
Then, we have 
\begin{align*}
    % \begin{aligned}
    %     \dot{\delta}_x =& (A_{11}-L_{1s}C_{1s} - L_{1f}C_{2s}) \delta_x + (A_{12} - L_{1f}C_{2f})\delta_z \\
    %                     & + B_1 e_{u_s} + L_{1s} (e_{y_s}+\hat v_1) + L_{1f}(e_f+\hat v_2) \\
    %                 %\eqqcolon& f_{\delta_x}(\delta_x,e_{y_s}, e_{u_s},\delta_z, e_{uf}, e_f)
    % \end{aligned}
    %\\
    \begin{aligned}
        \epsilon \dot{\delta}_z =& (A_{21}-L_{2s}C_{1s} - L_{2f}C_{2s}) \delta_x + (A_{22} - L_{2f}C_{2f})\delta_z \\
                                 & + B_2 e_{u_s} + L_{2s} (e_{y_s}+\hat v_1) + L_{2f}(e_f+\hat v_2)
                                 \\
                    \eqqcolon& g_{\delta_z}(\delta_x,e_{y_s}, e_{u_s},\delta_z, e_f,\hat{v}).
    \end{aligned} \label{eqn: delta_z dot}
\end{align*}
Let
\begin{equation}
    A_{11}^f\coloneqq A_{22}-L_{2f}C_{2f}.
    \label{eqn: A11f - Observer}
\end{equation}
Since the pair $(A_{11}^f, C_{2f})$ is observable, there exists a gain $L_{2f}$ such that the matrix $A_{11}^f - L_{2f}C_{2f}$ is Hurwitz. With this choice of $L_{2f}$, the fast subsystem is stabilized, thereby ensuring the existence of a well-defined quasi-steady state.
Let $\overline{\delta}_z$ and $\overline{e}_{f}$ denote the quasi-steady-state of $\delta_z$ and $e_f$, respectively. In particular, $\overline{e}_{f}=0$ due to sufficient transmissions, and $\overline{\delta}_z = \overline{H}(\delta_x,e_{u_s},e_{y_s},\hat{v})$, where 
\begin{multline*}
     \overline{H}(\delta_x,e_{u_s},e_{y_s},\hat{v})
    \coloneqq -{A_{11}^f}^{-1}  \big[ (A_{21}-L_{2s}C_{1s} - L_{2f}C_{2s}) \delta_x   \\
    + L_{2s}(e_{y_s}+\hat v_1)  +L_{2f} \hat v_2 + B_2 e_{u_s}  \big]
\end{multline*}
is the unique solution to $0=g_{\delta_z}(\delta_x,e_{y_s}, e_{u s},\overline{\delta}_z, 0,\hat{v})$, and $A_{11}^f$ is invertible as it is Hurwitz. 
We define the variable $\delta_y\coloneqq \delta_z - \overline{H}(\delta_x, e_{y_s},e_{u_s},\hat{v})$ that shifts the equilibrium of the fast state from its quasi-steady state to the origin, then at each slow transmission, $\delta_y$ is updated as follows:
\begin{equation}
    \begin{aligned}
        \delta_y^+ =& \delta_z^+ - \overline{H}(\delta_x^+, e_{u_s}^+, e_{y_s}^+,\hat{v}^+) \\
        % =& \delta_z - \overline{H}\big(\delta_x, h_{us}(\kappa_s, e_{u_s}), h_{ys}(\kappa_s, e_{y_s}, \tilde e_{p_s}),(h_{v_1}(\kappa_s, \hat v_1, v_1, \tilde e_{p_s}), \hat v_2)\big) \\
        % =& \delta_y + \overline{H}(\delta_x, e_{y_s}, e_{u_s},\hat{v}) 
        %     \\
        %     &-  \overline{H}\big(\delta_x, h_{us}(\kappa_s, e_{u_s}), h_{ys}(\kappa_s, e_{y_s}, \tilde e_{p_s}),(h_{v_1}(\kappa_s, \hat v_1, v_1, \tilde e_{p_s}), \hat v_2)\big) \\
        % \eqqcolon & h_y(\kappa_s, e_{y_s}, e_{u_s}, \tilde e_{p_s},\delta_y, \hat v_1, v_1). 
        =& \delta_y + \overline{H}(\delta_x, e_{y_s}, e_{u_s},\hat{v}) -  \overline{H}(\delta_x^+, e_{y_s}^+, e_{u_s}^+,\hat{v}^+) \\
        \eqqcolon & h_y(\kappa_s, e_{y_s}, e_{u_s}, \tilde e_{p_s},\delta_y, \hat v_1, v_1). 
    \end{aligned}
    \label{eqn: Jump of y at slow transmission - Observer}
\end{equation}
where $\overline{H}(\delta_x^+, e_{u_s}^+, e_{y_s}^+,\hat{v}^+) = \overline{H}\big(\delta_x, h_{us}(\kappa_s, e_{u_s}), h_{ys}(\kappa_s, e_{y_s}, $ $ \tilde e_{p_s}),(h_{v_1}(\kappa_s, \hat v_1, v_1, \tilde e_{p_s}), \hat v_2)\big)$.

\subsection{Full order system}
To formulate the SPNCS as a hybrid model, we define the full state vector \begin{equation*}\xi \coloneqq (\xi_s, \zeta_s,\xi_f,\zeta_f) \in \mathbb{X},
\end{equation*}
where
\begin{equation}
\begin{aligned}
   \xi_s &\coloneqq (\delta_x,e_{y_s},e_{u_s},\tau_s, \kappa_s, \hat v_1),\quad &\zeta_s \coloneqq (x_p, \tilde e_{p_s}), \\
   \xi_f & \coloneqq (\delta_y, e_f,\tau_f, \kappa_f, \hat v_2),\quad &\zeta_f \coloneqq (z_p,\tilde e_{p_f}),
\end{aligned}
    \label{eqn: definition of xi_s and xi_f - Observer}
\end{equation} 
with $e_{y_s}$ and $e_f$ defined in \eqref{eqn: ideal errors}, $e_{u_s}$, $\tilde e_{p_s}$, $\tilde e_{p_f}$ defined in \eqref{eqn: real plant network-induced errors}

Here, $\tau_s, \tau_f \in \mathbb{R}_{\geq 0}$ denote clocks that measure the elapsed time since the most recent slow and fast transmission, respectively, and $\kappa_s, \kappa_f \in \mathbb{Z}_{\geq 0}$ are counters for the number of slow and fast transmissions. The state space is defined as 
$\mathbb{X}\coloneqq \mathbb{R}^{n_x}\times \mathbb{R}^{n_{y_s}}\times \mathbb{R}^{n_{u_s}}  \times  \mathbb{R}_{\geq 0} \times \mathbb{Z}_{\geq 0}\times  \mathbb{R}^{n_{y_s}} \times \mathbb{R}^{n_x} \times  \mathbb{R}^{n_{y_s}}   \times\mathbb{R}^{n_z}\times \mathbb{R}^{n_{y_f}}\times \mathbb{R}_{\geq 0} \times \mathbb{Z}_{\geq 0}\times \mathbb{R}^{n_{y_f}} \times   \mathbb{R}^{n_z} \times \mathbb{R}^{n_{y_f}} $,
and the SPNCS is modeled by the hybrid system \citep{gosate12} with hybrid input $u_s$ and $v$, where $v$ is defined in \eqref{eqn: definition of v and v_hat}.
\begin{equation}
    \mathcal{H}:\left\{
\begin{aligned}
    \dot{\xi} &= F(\xi, \hat v, u_s, \dot{u}_s, \dot v, \epsilon),\ \xi \in \mathcal{C}^{\epsilon}, \\
    {\xi}^+ &\in G(\xi, v), \ \xi\in \mathcal{D}_s^{\epsilon} \cup \mathcal{D}_f^{\epsilon},
\end{aligned}
    \right.
    \label{eqn: H}
\end{equation}
where 
% \begin{equation*}
%     F(\xi,\hat{v}, u_s,\dot{u}_s, \dot v, \epsilon) = \begin{bmatrix}
%         F_{\xi_s}(\xi_s,\delta_y,e_f,\hat{v},\dot{u}_s) 
%         \\
%         F_{\zeta_s}(x_p,z_p,u_s,\dot v_1)
%         \\
%         \tfrac{1}{\epsilon}F_{\xi_f}(\xi,\hat{v},\dot{u}_s, \epsilon)
%         \\
%         \tfrac{1}{\epsilon} F_{\zeta_f}(x_p,z_p,u_s,\dot v_2,\epsilon)
%     \end{bmatrix},
% \end{equation*}
$F(\xi,\hat{v}, u_s,\dot{u}_s, \dot v, \epsilon) = \big(F_{\xi_s}(\xi_s,\delta_y,e_f,\hat{v},\dot{u}_s),  \allowbreak F_{\zeta_s}(x_p,\allowbreak z_p,u_s,\dot v_1),  \tfrac{1}{\epsilon}F_{\xi_f}(\xi,\hat{v},\dot{u}_s, \epsilon), \tfrac{1}{\epsilon} F_{\zeta_f}(x_p,z_p,u_s,\dot v_2,\epsilon)\big)$, 
and $F_{\xi_s}$ and $F_{\xi_f}$ are defined in \eqref{eqn: Functions}. Moreover, $F_{\zeta_s}(x_p,z_p,u_s, \dot v_1) = (\dot{x}_p,-C_{1s}\dot{x}_p - \dot{v}_1)$ and $F_{\zeta_f}(x_p,z_p,u_s,\dot v_2,\epsilon) = (\dot{z}_p, -\epsilon C_{2s}\dot{x}_p - C_{2f}\dot{z}_p - \epsilon \dot{v}_2) $.
The jump map $G$ is given by
\begin{equation}
\begin{aligned}
    G(\xi) \coloneqq \left\{ 
    \begin{aligned}
    &G_s(\xi, v_1),  \;\xi\in\mathcal{D}_s^{\epsilon} \setminus \mathcal{D}_f^{\epsilon} , \\
    &G_f(\xi, v_2),  \;  \xi \in\mathcal{D}_f^{\epsilon} \setminus \mathcal{D}_s^{\epsilon} ,\\
    &\{G_s(\xi,v_1),G_f(\xi,v_2)\}, \; \xi\in \mathcal{D}_s^{\epsilon} \cap \mathcal{D}_f^{\epsilon},
    \end{aligned}
    \right. 
\end{aligned}
\label{eqn: G - Observer}
\end{equation}
where $G_s(\xi,v_1) = (G_{s,1}(\xi_s, \zeta_s,v_1),G_{s,2}(\xi, v_2))$, and $G_{s,1}$, $G_{s,2}$ and $G_f$ are defined in \eqref{eqn: Functions}. The jump maps $G_s$ and $G_f$ correspond to the transmissions of slow and fast signals, respectively. 
Let $\tau_{\text{mati}}^f = \epsilon T^*$ with $T^* \in \mathbb{R}_{>0}$ independent of $\epsilon$ and define $\tau_{\text{miati}}^f = a\tau_{\text{mati}}^f$ for some $a \in(0,\tfrac{1}{2}] $, satisfying the inequality $\tau_{\text{miati}}^f \leq  \tfrac{1}{2}\tau_{\text{mati}}^f$. Since $\epsilon > 0$, the condition $\epsilon \tau_f \in [\tau_{\text{miati}}^f, \tau_{\text{mati}}^f]$ is equivalent to $\tau_f \in [aT^*,T^*]$. Consequently, the jump and flow sets in \eqref{eqn: H} are defined as
\begin{equation*}
    \begin{aligned}
        \mathcal{D}_s^{\epsilon} \! &\coloneqq \!  \{\xi \in \mathbb{X} \; | \; \tau_s \in [\tau_{\text{miati}}^s, \tau_{\text{mati}}^s] \wedge \tau_f \in  [aT^*, (1-a)T^*] \}, \\
        \mathcal{D}_f^{\epsilon} \! & \coloneqq \! \{\xi \in \mathbb{X} \; | \; \tau_s \in [\epsilon aT^*, \tau_{\text{mati}}^s-\epsilon aT^*]  \wedge  \tau_f \in  [aT^*, T^*]  \}.
    \end{aligned}
\end{equation*}
% $\mathcal{D}_s^{\epsilon} \! \coloneqq \!  \{\xi \in \mathbb{X} \; | \; \tau_s \in [\tau_{\text{miati}}^s, \tau_{\text{mati}}^s] \wedge \tau_f \in  [aT^*, (1-a)T^*] \}$ and $\mathcal{D}_f^{\epsilon} \! \coloneqq \! \{\xi \in \mathbb{X} \; | \; \tau_s \in [\epsilon aT^*, \tau_{\text{mati}}^s-\epsilon aT^*]  \wedge  \tau_f \in  [aT^*, T^*]  \}$.
% \begin{align*}
%     \mathcal{D}_s^{\epsilon} \! \coloneqq \!  \{\xi \in \mathbb{X} \; | \; \tau_s \in [\tau_{\text{miati}}^s, \tau_{\text{mati}}^s] \wedge \tau_f \in  [aT^*, (1-a)T^*] \},
%     \\
%     \mathcal{D}_f^{\epsilon} \! \coloneqq \! \{\xi \in \mathbb{X} \; | \; \tau_s \in [\epsilon aT^*, \tau_{\text{mati}}^s-\epsilon aT^*]  \wedge  \tau_f \in  [aT^*, T^*]  \}.
% \end{align*}
This construction of $G$ guarantees the outer semi-continuity and the local boundness of the jump map, which is one of the hybrid basic conditions \citep{gosate12}.
The flow set is given by $\mathcal{C}^{\epsilon} \coloneqq 
        \mathcal{D}_s^{\epsilon} \cup \mathcal{D}_f^{\epsilon} \cup \mathcal{C}_{a}^{\epsilon} \cup \mathcal{C}_{b}^{\epsilon}$,
with 
% \begin{align*}
%     \mathcal{C}_{a}^{\epsilon} &\coloneqq \{ \xi \in \mathbb{X} \ | \ \tau_s \in [0, \epsilon a T^*]  \wedge \epsilon \tau_f \in [0,\tau_s + \epsilon T^* - \epsilon a T^*] \},
%     \\
%     \mathcal{C}_{b}^{\epsilon} &\coloneqq \{ \xi \in \mathbb{X} \;|\; \tau_s \in [\epsilon a T^*, \epsilon \tau_f + \tau_{\text{mati}}^s - \epsilon a T^*]  \wedge \epsilon \tau_f \in  [0, \epsilon a T^*]  \}.
% \end{align*}
$\mathcal{C}_{a}^{\epsilon} \coloneqq \{ \xi \in \mathbb{X} \ | \ \tau_s \in [0, \epsilon a T^*]  \wedge \epsilon \tau_f \in [0,\tau_s + \epsilon T^* - \epsilon a T^*] \} $ and 
$\mathcal{C}_{b}^{\epsilon} \coloneqq \{ \xi \in \mathbb{X} \;|\; \tau_s \in [\epsilon a T^*, \epsilon \tau_f + \tau_{\text{mati}}^s - \epsilon a T^*]  \wedge \epsilon \tau_f \in  [0, \epsilon a T^*]  \}$.
\begin{remark}
    The states $\zeta_s$ and $\zeta_f$ do not affect the dynamics of estimation errors during flow; they appear only in the jump map of $e_{y_s}$, $e_f$ and $\hat v$. If a static protocol such as round-robin is implemented, $\zeta_s$ and $\zeta_f$ decouple from the estimation errors and their dynamics need not be included in $\mathcal{H}$. 
\end{remark}
%We are now ready to derive the reduced system and boundary-layer system associated with $\mathcal{H}$.

\subsection{Boundary-layer system and reduced system}
 We first derive the boundary-layer system. Define the fast time scale as $\sigma \coloneqq \tfrac{t}{\epsilon}$,
 so that $\tfrac{\partial}{\partial \sigma} = \epsilon \tfrac{\partial}{\partial t}$. By setting $\epsilon = 0$ in system \eqref{eqn: H}, the corresponding flow set $\mathcal{C}^{0}$ is given by $$\mathcal{C}^{0} \coloneqq \{ \xi \in \mathbb{X} \ | \ \tau_s \in [0, \tau_\text{mati}^s]  \wedge \tau_f \in [0,  T^*] \}  ,$$ with $\mathcal{D}_s^{0}$, $\mathcal{D}_f^{0}$ derived accordingly. In the fast subsystem, the slow dynamics are effectively frozen, and the condition $\tau_{s}\in [0, \tau_{\text{mati}}^s]$ is always satisfied. Thus, the evolution of the boundary-layer system $\mathcal{H}_{bl}$ is governed solely by $\tau_{f}$. Consequently, we write
\begin{equation}
    \mathcal{H}_{bl}\! : \! \left\{
\begin{aligned}
    \tfrac{\partial \xi}{\partial \sigma}  &= (0,0, F_{\xi_f}(\xi_s, \xi_f,\hat{v},\dot{u}_s, 0),  \\
                                            & \qquad \quad  F_{\zeta_f}(x_p,z_p,u_s,0) ), \quad \xi \! \in \mathcal{C}_{bl}^{0}, \\
    {\xi}^+  &=   G_f(\xi, v_2), \qquad \qquad  \qquad  \, \xi \! \in \mathcal{D}_f^{0},
\end{aligned}
    \right.
    \label{eqn: H_bl - Observer}
\end{equation}
where $\mathcal{C}_{bl}^{0} \coloneqq \{\xi \in  \mathbb{X} \ | \ \tau_f \in [0, T^*]\}$, $\mathcal{D}_f^{0}\coloneqq \{\xi \in  \mathbb{X} \ | \ \tau_f \in [aT^*, T^*] \}$, $F_{\xi_f}$ and $G_f$ are defined in \eqref{eqn: Functions}, and $F_{\zeta_f}$ is defined under \eqref{eqn: H}.
For clarity in stating our assumptions, we explicitly state the dynamic component of the boundary-layer system as
$$
(\tfrac{\partial \delta_y}{\partial \sigma}, \tfrac{\partial {e_f}}{\partial \sigma})
    =
    \begin{bmatrix}
        A_{11}^f & A_{12}^f \\ A_{21}^f & A_{22}^f
    \end{bmatrix} 
    \begin{bmatrix}
        \delta_y \\ e_f
    \end{bmatrix} ,
$$
% \begin{equation*}
%     \left[\begin{smallmatrix}
%         \tfrac{\partial \delta_y}{\partial \sigma} \\ \tfrac{\partial {e_f}}{\partial \sigma}
%     \end{smallmatrix} \right]
%     =
%     \left[\begin{smallmatrix}
%         A_{11}^f & A_{12}^f \\ A_{21}^f & A_{22}^f
%     \end{smallmatrix} \right]
%     \left[\begin{smallmatrix}
%         \delta_y \\ e_f
%     \end{smallmatrix} \right],
% \end{equation*}
where $A_{11}^f $ is defined in \eqref{eqn: A11f - Observer}, $A_{12}^f \coloneqq L_{2f}$, $A_{21}^f \coloneqq C_{2f}A_{11}^f$ and $A_{22}^f \coloneqq C_{2f} L_{2f}$. %\red{We note that, Assumption \ref{Assumption A11f - Observer} implies that the observer gain $L_{2f}$ is selected such that the boundary-layer system $H_{bl}$ is UGES under perfect network (i.e., $e_f = 0$).}

From the perspective of slow dynamics, the fast dynamics evolve infinitely fast and thus remain at their quasi-steady state. %(i.e., $\delta_y=0$ and $e_f = 0$). 
Consequently, the reduced system $\mathcal{H}_r$ is given by
\begin{equation}
    \mathcal{H}_r  \! \coloneqq \!\left\{
\begin{aligned}
    (\dot \xi_s, \dot \zeta_s) &= (F_{\xi_s}(\xi_s,0,0,\hat{v},\dot{u}_s) , \\
    & \qquad F_{\zeta_s}(x_p,\overline{z}_p,u_s,\dot v_1)), \quad  \xi \in \mathcal{C}_{r}^{0}, \\
    (\xi_s^+,\zeta_s^+)  &=   G_{s,1}(\xi_s, \zeta_s,v_1), \qquad \quad  \xi\in \mathcal{D}_s^{0},
\end{aligned}
    \right.
    \label{eqn: H_r - Observer}
\end{equation}
where $\overline{z}_p \coloneqq -A_{22}^{-1} (A_{21}x_p + B_2 u_s)$ is the quasi-steady-state of $z_p$, $\mathcal{C}_{r}^{0} \coloneqq \{\xi \in  \mathbb{X} \ | \ \tau_s \in [0, \tau_{\text{mati}}^s]\}$ and $\mathcal{D}_s^{0}\coloneqq \{\xi \in  \mathbb{X} \ | \ \tau_s \in [\tau_{\text{miati}}^s, \tau_{\text{mati}}^s] \}$.
%
%
% We write part of the dynamic of reduced system as following:
% \begin{equation*}
%     \left[ \begin{smallmatrix}
%         \dot \delta_x \\ \dot e_{y_s} \\ \dot e_{u_s}
%     \end{smallmatrix} \right]
%     =
%     \left[ \begin{smallmatrix}
%         A_{11}^r & A_{12}^r & A_{13}^r \\ A_{21}^r & A_{22}^r & A_{23}^r \\ 0&  0& 0
%     \end{smallmatrix} \right]
%     %
%     \left[ \begin{smallmatrix}
%      \delta_x \\  e_{y_s} \\  e_{u_s}
%     \end{smallmatrix} \right]
%     +
%     \left[ \begin{smallmatrix}
%         A_{14}^r & A_{15}^r  \\ A_{24}^r & A_{24}^r
%     \end{smallmatrix} \right]  
%     %
%     \left[ \begin{smallmatrix}
%         \hat v_1 \\ \hat v_2
%     \end{smallmatrix} \right]
%     +
%     \left[ \begin{smallmatrix}
%         0\\ - \dot \hat v_1 \\ - \dot u_s
%     \end{smallmatrix} \right]
% \end{equation*}
% where $A_{11}^r = (A_{11}-L_{1s}C_{1s} - L_{1f}C_{2s}) - D (A_{21}- L_{2s}C_{1s}-L_{2f}C_{2s})  $,
% $A_{12}^r = L_{1s}- D L_{2s}$,
% $A_{13}^r = B_1- D B_2$,
% $A_{14}^r = A_{12}^r$,
% $A_{15}^r = L_{1f}- D L_{2f}$, $A_{2 \star}^r = C_{1s}A_{1\star}^r$ for all $\star \in \{ 1,\cdots,5\}$ and $D = (A_{12} - L_{1f}C_{2f}) (A_{22} - L_{2f}C_{2f})^{-1}  $.
%
Define the error vector and its derivative
\begin{equation}
    e_s \coloneqq (e_{y_s}, e_{u_s}) \text{, }\dot e_s = f_{es}(\delta_x,e_{y_s}, e_{u_s},\delta_y, e_f,\hat{v},\dot u_s).
    \label{eqn: definition e_s and f_es}
\end{equation}
%and let its derivative be denoted by $\dot e_s = f_{es}(\delta_x,e_{y_s}, e_{u_s},\delta_y, e_f,\hat{v},\dot u_s)$. 
Furthermore, we denote its update rule at slow transmission as $$h_s(\kappa_s, e_s, \tilde e_{p_s}) \coloneqq \big(h_{ys}(\kappa_s, e_{y_s},\tilde e_{p_s}), h_{u_s}(\kappa_s, e_{u_s}) \big).$$
Then, a portion of the dynamics of the reduced system is given by
\begin{equation*}
    \left[ \begin{smallmatrix}
        \dot \delta_x \\ \dot e_{s} 
    \end{smallmatrix} \right]
    =
    \left[ \begin{smallmatrix}
        A_{11}^s & A_{12}^s  \\ A_{21}^s & A_{22}^s
    \end{smallmatrix} \right]
    \left[ \begin{smallmatrix}
     \delta_x \\  e_{s} 
    \end{smallmatrix} \right]
    +
    \left[ \begin{smallmatrix}
        A_{13}^s \\ A_{23}^s
    \end{smallmatrix} \right]  
    \left[ \begin{smallmatrix}
        \hat v_1 \\ \hat v_2
    \end{smallmatrix} \right]
    +
    (\mathbf{0}_{n_x\times 1}, \left[ \begin{smallmatrix}
        \mathbf{0}_{n_{y_s}}\times 1 \\ - \dot u_s
    \end{smallmatrix} \right]),
\end{equation*}
% $
%     \left[ \begin{smallmatrix}
%         \dot \delta_x \\ \dot e_{s} 
%     \end{smallmatrix} \right]
%     =
%     \left[ \begin{smallmatrix}
%         A_{11}^s & A_{12}^s  \\ A_{21}^s & A_{22}^s
%     \end{smallmatrix} \right]
%     %
%     \left[ \begin{smallmatrix}
%      \delta_x \\  e_{s} 
%     \end{smallmatrix} \right]
%     +
%     \left[ \begin{smallmatrix}
%         A_{13}^s \\ A_{23}^s
%     \end{smallmatrix} \right]  
%     %
%     \left[ \begin{smallmatrix}
%         \hat v_1 \\ \hat v_2
%     \end{smallmatrix} \right]
%     +
%     (\mathbf{0}_{n_x\times 1}, \left[ \begin{smallmatrix}
%         \mathbf{0}_{n_{y_s}}\times 1 \\ - \dot u_s
%     \end{smallmatrix} \right]),
% $
where $A_{11}^s \coloneqq A_{11}^r$, 
$A_{12}^s \coloneqq \left[\begin{smallmatrix} A_{12}^r & A_{13}^r\end{smallmatrix} \right]$,
$A_{13}^s \coloneqq \left[\begin{smallmatrix}A_{14}^r & A_{15}^r \end{smallmatrix} \right]$,
$A_{21}^s \coloneqq \left[\begin{smallmatrix} A_{21}^r & \mathbf{0}\end{smallmatrix} \right]^\top$, 
$A_{22}^s \coloneqq \left[\begin{smallmatrix}A_{22}^r & A_{23}^r \\ \mathbf{0} & \mathbf{0} \end{smallmatrix} \right]$,
$A_{23}^s \coloneqq \left[\begin{smallmatrix}A_{24}^r & A_{25}^r \\ \mathbf{0} & \mathbf{0} \end{smallmatrix} \right]$,
$A_{11}^r \coloneqq (A_{11}-L_{1s}C_{1s} - L_{1f}C_{2s}) - D (A_{21}- L_{2s}C_{1s}-L_{2f}C_{2s})  $,
$A_{12}^r \coloneqq L_{1s}- D L_{2s}$,
$A_{13}^r \coloneqq B_1- D B_2$,
$A_{14}^r \coloneqq A_{12}^r$,
$A_{15}^r \coloneqq L_{1f}- D L_{2f}$, $A_{2 \star}^r \coloneqq C_{1s}A_{1\star}^r$ for all $\star \in \{ 1,\cdots,5\}$ and $D \coloneqq (A_{12} - L_{1f}C_{2f}) (A_{22} - L_{2f}C_{2f})^{-1}  $.

\section{Stability analysis}%
\label{Section Analysis - Observer}
We begin by assuming that the scheduling protocol $h_\star$ is UGES with the associate Lyapunov function, as suggested in \cite[Definition 7]{dragan_stability}. Let $n_{e_s} = n_{y_s}+n_{u_s}$ and $n_{e_f} = n_{y_f}$, we propose the following assumption on our scheduling protocol.
\begin{ass}
    For all $\star \in \{s,f\}$, there exist a function $W_\star: \mathbb{Z}_{\geq 0}\times \mathbb{R}^{n_{e_\star}} \to \mathbb{R}_{\geq 0}$ that is locally Lipschitz in its second argument.
    Moreover, there exist constants $\underline{a}_{W_\star},\overline{a}_{W_\star} >0 $ as well as a constant $\lambda_\star \in [0,1)$ such that, for all $ \kappa_\star \in \mathbb{Z}_{\geq 0}$ and $e_\star \in \mathbb{R}^{n_{e_\star}}$, the following properties hold:
\begin{align}
    \underline{a}_{W_\star}\left|  {e_\star}  \right| \leq {W_\star}(k_\star, {e_\star}) \leq \overline{a}_{W_\star}\left|   {e_\star}  \right| ,\label{eqn: W sandwich bound}
    \\
    {W_\star}(\kappa_\star + 1, h_\star(\kappa_\star, e_\star, 
    \tilde{e}_{p_\star})) \leq \lambda_\star {W_\star}(\kappa_\star, {e_\star}). \label{eqn: W jump}
\end{align}
Moreover, there exists $M_\star \geq 0$, such that for all $\kappa_\star \in \mathbb{Z}_{\geq 0}$ and almost all ${e_\star} \in \mathbb{R}^{n_{e_\star}}$, $\left| \tfrac{\partial W_\star(\kappa_\star,e_\star)}{\partial e_\star} \right| \leq M_\star $.
\label{Assumption UGES protocol}
\end{ass} 
By definition of $f_{es}$ in \eqref{eqn: definition e_s and f_es}, there exist $A_{H_s} = M_s A_{21}^s$ and  $L_s  = M_s |A_{22}^s| \underline{a}_{W_s}^{-1}$, such that for all $\delta_x \in \mathbb{R}^{n_x}, \kappa_s \in \mathbb{Z}_{\geq 0}$ and almost all ${e_s} \in \mathbb{R}^{n_{e_s}}$, the following inequality holds,
\begin{multline}    
    \left< \tfrac{\partial {W_s}(\kappa_s,{e_s})}{\partial {e_s}}, f_{es}(\delta_x,e_{ys}, e_{us},0, 0,\hat v, \dot u_s)\right> 
    \leq  \\ L_s {W_s}(\kappa_s, e_s)  + |A_{H_s} \delta_x| + \sigma_{w_s}(|\hat v_1|,|\hat v_2|, |\dot u_s|),
    \label{eqn: reduced system Ws dot}
\end{multline}
where 
\begin{equation}
    \sigma_{w_s}(|\hat v_1|,|\hat v_2|, |\dot u_s|) \coloneqq a_{w_s}^{v_1} |\hat v_1| + a_{w_s}^{v_2} |\hat v_2| + M_s |\dot u_s|, \label{eqn: sigma_Ws}
\end{equation}
 $a_{w_s}^{v_1} = M_s\left|A_{24}^r  \right|$ and $a_{w_s}^{v_2} = M_s\left|A_{25}^r  \right|$.
This inequality bounds the growth of network-induced error during the flow of the reduced system.

Similarly, by definition of $g_{ef}$ in \eqref{eqn: Functions}, there exist $ L_f = M_f |a_{22}^f| \underline{a}_{W_f}^{-1}$ and $A_{H_f} = M_f A_{21}^f$ such that for all $x \in \mathbb{R}^{n_x}, \kappa_f \in \mathbb{Z}_{\geq 0}$, and almost all ${e_{f}} \in \mathbb{R}^{n_{y_f}}$, the following inequality holds:
\begin{multline}
   \left< \tfrac{\partial {W_f}(\kappa_f,{e_{f}})}{\partial {e_{f}}}, g_{e_f}(\delta_x,e_{ys}, e_{us},\delta_y, e_f, \hat v, 0)\right>
     \leq  \\ L_f {W_f}(\kappa_f, e_{f}) + |A_{H_f}\delta_y|.
    \label{eqn: boundary-layer system Wf dot}
\end{multline}

%Next, we introduce Assumptions \ref{Assumption boundary-layer system - Observer} and \ref{Assumption reduced model - Observer} to ensure the robustness of $\mathcal{H}_r$ and $\mathcal{H}_{bl}$, respectively. 

The following two assumptions guide a design procedure aimed at minimizing the gain between the network-induced error and the estimation error from the observer. A smaller gain enhances robustness to network-induced errors, allowing for slower transmission without compromising performance. Another approach is to directly maximize the transmission interval, as demonstrated in our numerical example.
Let $P_s$, $P_f$ be positive definite symmetric matrices, and $a_{\rho_s}$, $\gamma_s$, $ \eta_1$, $ a_{\rho_f}$, $\gamma_f >0$. We introduce the following design condition, which is instrumental in reducing conservatism in the resulting MATI.
\begin{con}
The variables $L_{2f}$, $P_f$ and $a_{\rho_f}$ are chosen so that \eqref{eqn: LMI BL} holds with \textbf{the minimum feasible value of $\gamma_f$}.
\begin{equation}
       \left[\begin{smallmatrix}
        P_f A_{11}^f + A_{11}^{f\top} P_f + a_{\rho_f}I + A_{H_f}^\top A_{H_f}& P_f A_{12}^f \\ A_{12}^{f\top} P_f & a_{\rho_f}\overline{a}_{W_f}^2 I - \gamma_f^2 \underline{a}_{W_f}^2 I
    \end{smallmatrix} \right] \leq 0. \label{eqn: LMI BL}
\end{equation}
\label{Assumption boundary-layer system - Observer}
\end{con}%   
The LMI \eqref{eqn: LMI BL} is always feasible 
since the pair $(A_{22}, C_{2f})$ is observable. In particular, one can select $L_{2f}$ such that $A_{11}^f$ is Hurwitz, then there always exist $P_f > 0$ such that 
\begin{equation}
P_f A_{11}^f + A_{11}^{f\top} P_f + a_{\rho_f}I + A_{H_f}^\top A_{H_f} < 0.
\label{eqn: Schur condition 1}
\end{equation}
Moreover, by \eqref{eqn: Schur condition 1}, we have
\begin{equation*}
(P_f A_{11}^f + A_{11}^{f\top} P_f + a_{\rho_f}I + A_{H_f}^\top A_{H_f})^{-1} < 0,
\end{equation*}
which also implies 
\begin{equation*}
    A_{12}^{f\top} P_f(A_{11}^f + A_{11}^{f\top} P_f + a_{\rho_f}I + A_{H_f}^\top A_{H_f}P_f)^{-1} P_f A_{12}^f \leq 0.
\end{equation*}
Since we can always find sufficiently large $\gamma_f$ such that 
\begin{multline*}
        (a_{\rho_f}\overline{a}_{W_f}^2 I - \gamma_f^2 \underline{a}_{W_f}^2 I)  
        \\
        -A_{12}^{f\top} P_f(A_{11}^f + A_{11}^{f\top} P_f + a_{\rho_f}I + A_{H_f}^\top A_{H_f}P_f)^{-1} P_f A_{12}^f \leq 0,
\end{multline*}
the LMI \eqref{eqn: LMI BL} is satisfied by Schur complement.

\begin{con}
Let $L_{2f}$ comes from Assumption \ref{Assumption boundary-layer system - Observer}. Then, $L_{1s}$, $L_{1f}$, $L_{2s}$, $P_s$, $a_{\rho_s}$ and $\eta_1$ are chosen such that \eqref{eqn: LMI reduced} holds with \textbf{the minimum feasible value of $\gamma_s$}.
\begin{equation}
    \left[\begin{smallmatrix}
        P_s A_{11}^s + A_{11}^{s\top} P_s + \eta_1 P_s^\top P_s + a_{\rho_s}I + A_{H_s}^\top A_{H_s}& P_s A_{12}^s \\ A_{12}^{s\top} P_s & a_{\rho_s}\overline{a}_{W_s}^2I - \gamma_s^2 \underline{a}_{W_s}^2 I
    \end{smallmatrix} \right] \leq 0.
    \label{eqn: LMI reduced}
\end{equation}
\label{Assumption reduced model - Observer}
\end{con}
The feasibility of \eqref{eqn: LMI reduced} can be checked along the same line as \eqref{eqn: LMI BL}.
The proof of Theorem \ref{Theorem H - Observer} will show that inequality \eqref{eqn: LMI reduced} implies $\mathcal{L}_2$ stability of reduced system \eqref{eqn: H_r - Observer} from $W_s(\kappa_s, e_s)$ to $|A_{Hs}\delta_x|$ with gain $\gamma_s$, and input-to-state stability of \eqref{eqn: H_r - Observer} w.r.t. $\hat v$. 
Similarly, \eqref{eqn: LMI BL} guarantee the $\mathcal{L}_2$ stability of the boundary-layer system \eqref{eqn: H_bl - Observer}. 

% \begin{remark}
%     \cyan{
%     Assumptions \ref{Assumption boundary-layer system - Observer} and \ref{Assumption reduced model - Observer} imply that $A_{11}^s$ and $A_{11}^f$ are Hurwitz, which guarantee the stability of the continuous-time error dynamics for sufficiently small $\epsilon$ \cite{nonlinear_systems_Khalil}. 
%     A smaller $\gamma_\star$ enhance robustness on network-induced error $e_\star$, leading to larger $\tau_{\text{mati}}^\star$. 
%     %Additionally, a larger $\eta_1$ reduces the gain related with $v$. 
%     Our illustrative example further improve Assumptions \ref{Assumption boundary-layer system - Observer} and \ref{Assumption reduced model - Observer} by directly maximizing $\tau_{\text{mati}}^f$ using LMIs, instead of minimizing $\gamma_s$ and $\gamma_f$.
%     }
% \end{remark}
% \vspace{-3mm}

%
Before stating our main result, we introduce the function $T(\cdot,\cdot,\cdot)$ \citep{carnevale_stability}, which provides an upper bound of $\tau_{\text{mati}}^s$ and $T^*$. Let
% \begin{equation}
%     T(L, \gamma, \lambda) \! \coloneqq \! 
%     \begin{cases}
%     \tfrac{1}{Lr}\tan^{-1} \! \! \bigg(\!\tfrac{r(1-\lambda)}{2\tfrac{\lambda}{1+\lambda}\big(\tfrac{\gamma}{L}-1\big)+1+\lambda} \!\bigg), \;  \gamma > L
%     \\
%     \tfrac{1}{L} \tfrac{1-\lambda}{1+\lambda}, \quad \qquad\qquad\qquad\qquad \;\;\; \gamma = L
%     \\
%     \tfrac{1}{Lr}\tanh^{-1} \! \! \bigg(\!  \tfrac{r(1-\lambda)}{2\tfrac{\lambda}{1+\lambda}\left(\tfrac{\gamma}{L}-1\right)+1+\lambda} \!  \bigg), \; \gamma < L ,
%     \end{cases}
%     \label{eqn: function T}
% \end{equation}
\begin{equation}
    T(L, \gamma, \lambda)  \coloneqq 
    \begin{cases}
    \tfrac{1}{Lr_1}\tan^{-1} (r_2), \quad  \gamma > L
    \\
    \tfrac{1}{L} \tfrac{1-\lambda}{1+\lambda}, \quad \qquad \;\;\;\;\; \;\; \gamma = L
    \\
    \tfrac{1}{Lr_1}\tanh^{-1} (r_2), \;\;\;\; \gamma < L ,
    \end{cases}
    \label{eqn: function T - Observer}
\end{equation}
where $r_1 \coloneqq \sqrt{|\left(\tfrac{\gamma}{L}\right)^2 -1|}$, $r_2 \coloneqq \tfrac{r_1(1-\lambda)}{2\tfrac{\lambda}{1+\lambda}\big(\tfrac{\gamma}{L}-1\big)+1+\lambda}$. In the special case where $L = 0$, this nonlinear mapping simplifies to
$T(0,\gamma, \lambda) = \tfrac{1}{\gamma} \big(\tan^{-1}(\tfrac{1}{\lambda}) - \tan^{-1}(\lambda) \big).$
Lastly, we define the following attractor $$\mathcal{E} \coloneqq \left\{ \xi \in \mathbb{X} : \delta_x=0 \wedge e_s = 0 \wedge \delta_y=0 \wedge e_f = 0 \right\}$$ for the upcoming stability analysis.
At this point, all necessary prerequisites have been established. We now state our main result.
\begin{theorem}
Consider the system \eqref{eqn: H} satisfying Assumptions~\ref{Assumption UGES protocol}, and Condition~\ref{Assumption boundary-layer system - Observer} and \ref{Assumption reduced model - Observer}. 
Let $L_s$ and $L_f$ come from \eqref{eqn: reduced system Ws dot} and \eqref{eqn: boundary-layer system Wf dot}, respectively, $\gamma_f$ and $\gamma_s$ come from Conditions \ref{Assumption boundary-layer system - Observer} and  \ref{Assumption reduced model - Observer}, respectively. Moreover, let $\lambda_s$ and $\lambda_f$ come from Assumption \ref{Assumption UGES protocol}, and $\lambda_f^*\in [\lambda_f,1)$.
Then, for any $\tau_{\text{miati}}^s \leq \tau_{\text{mati}}^s < T(L_s, \gamma_s, \lambda_s)$ and $\tau_{\text{mati}}^f \leq \epsilon T(L_f, \gamma_f,\lambda_f^*)$, the following holds:

There exist $k, a, \epsilon^* >0 $ and $\gamma_{v_1},\gamma_{v_2}, \gamma_{\dot u_s} \geq 0$ such that for all $\epsilon \in (0, \epsilon^* ]$, any solution $\xi$ to $\mathcal{H}$ with hybrid inputs $u_s$, $v_1$ and $v_2$, where $u_s$ is absolutely continuous and $\dot{u}_s, v_1, v_2 \in \mathcal{L}_\infty$, satisfies a DISS property, i.e., 
\begin{multline}
    |\xi(t,j)|_\mathcal{E} \leq k |\xi(0,0)|_\mathcal{E} e^{-a(t+j)} + \gamma_{v_1} \|v_1 \|_{(t,j)} 
    \\
    +\gamma_{v_2} \|v_2 \|_{(t,j)}+\gamma_{\dot u_s} \|\dot u_s \|_{(t,j)},
    \label{eqn: DISS}
\end{multline}
for any $(t,j)\in \text{dom}(\xi)$.
\label{Theorem H - Observer}
\end{theorem}
\textbf{Proof:} See Appendix \ref{Section: Proof of theorem observer}.
\begin{remark} \label{Remark: theorem - DISS}
    It is important to note that our result, presented in Theorem \ref{Theorem H - Observer}, shows that the ultimate bound depends on the derivative of the input signal and the magnitude of the measurement noise. This result is a special case of the standard DISS introduced in \cite{DISS}.
\end{remark}
\begin{remark} \label{Remark: theorem - Observer}
    A fundamental tradeoff exists between the MATI estimates and the ultimate bounds (i.e., $\gamma_{v_1}$, $\gamma_{v_2}$, and $\gamma_{\dot u_s}$). In particular, achieving larger MATI estimates typically results in increased ultimate bounds. This reflects the fact that allowing sparser transmissions reduces communication demands but at the expense of degraded robustness, as the network-induced errors and disturbance effects accumulate more significantly between transmissions.
\end{remark}

\section{Illustrative example}
\label{Section Example - Observer}
In this section, we provide a numerical example to illustrate the procedure for deriving the MATI and the corresponding ultimate bounds. We also present a simulation that demonstrates how variations in the control input influence the error dynamics, thereby highlighting the practical implications of the theoretical results.

Consider plant \eqref{eqn: plant} with
$A_{11} = \left[ \begin{smallmatrix}
    a_1 & 0 \\ 0 & -a_5
\end{smallmatrix} \right]$, 
$A_{12} = \left[ \begin{smallmatrix}
    a_2 & 0 \\ 0 & 0
\end{smallmatrix} \right]$,
$A_{21} = \left[ \begin{smallmatrix}
    0 & 0 \\ a_3 & 0
\end{smallmatrix} \right]$,
$A_{22} = \left[ \begin{smallmatrix}
    -a_2 & 0 \\ -a_2 & -a_4
\end{smallmatrix} \right]$,
$C_{1s} = \left[ \begin{smallmatrix}
    c_1 & c_2
\end{smallmatrix} \right]$,
$C_{2s} = \left[ \begin{smallmatrix}
    -1 & 0
\end{smallmatrix} \right]$,
$C_{2f} = \left[ \begin{smallmatrix}
    0 & -1
\end{smallmatrix} \right]$, where
$a_1 =10^{-3}$, $a_2 = 0.37$, $a_3 = 1.1$, $a_4 = 4.9$, $a_5 = 1.2$, $c_1 = -0.03$ and $c_2 = 1.9$.
To simplify the computation, we assume observer gains have the form 
$L_{1s} = \left[ \begin{smallmatrix}
    n_1 \\ 0
\end{smallmatrix} \right]$,
$L_{2s} = \left[ \begin{smallmatrix}
    -n_2 \\ -n_2
\end{smallmatrix} \right]$,
$L_{1f} = \left[ \begin{smallmatrix}
    0 \\ n_3
\end{smallmatrix} \right]$,
and
$L_{2f} = \left[ \begin{smallmatrix}
    0 \\ 0
\end{smallmatrix} \right]$.

Since $u_s$, $y_{p_s}$ and $y_{p_f}$ are scalars, the protocols are given by $h_s(\kappa_s, e_s) = 0$ and $h_f(\kappa_f, e_f) = 0$, which are UGES protocols. Define $W_\star(\kappa_\star,e_\star) \coloneqq |e_\star|$, $s\in \{s,f\}$. Then, Assumption \ref{Assumption UGES protocol} holds with $\underline{a}_{W_\star}(s) = \overline{a}_{W_\star}(s) = s$, $\lambda_\star = 0 $ and $M_\star = 1$. Moreover, inequality \eqref{eqn: reduced system Ws dot} is satisfied with $L_s = 1.85$ and $A_{H_s} = A_{21}^s$, and \eqref{eqn: boundary-layer system Wf dot} holds with $L_f = 0$ and $A_{H_f} = A_{21}^f$.

By solving the LMI \eqref{eqn: LMI BL} and \eqref{eqn: LMI reduced} simultaneously while maximizing $ \epsilon^* T^*$, we obtain the parameters in observer gains to be  $n_1 = 0.02$, $n_2 = 0.01$ and $n_3 = 0.18$. Meanwhile, we have
$P_s = \left[\begin{smallmatrix} 0.14  & -0.55\\ -0.55 & 2.8\end{smallmatrix} \right]$, $\gamma_s = 3.4$, $a_{\rho_s} = 0.14$, $P_f = \left[\begin{smallmatrix} 3.5  & 0.15\\ 0.15 & 2.7\end{smallmatrix} \right]$ and $\gamma_f = 1.6$, $a_{\rho_f} = 2.5$.
We select $\tau_{\text{mati}}^s = 0.15 \ s < T(L_s,\gamma_s, \lambda_s)$. Then, with $\lambda_s^* = 0.33$, $\eta_{s1} = 0.1$, $\eta_{s2} = 0.1$, $\eta_{s3} = 0.5$ and $ \eta_s = \eta_{s1}+ \eta_{s2} + \eta_{s3} =0.7$, Lemma \ref{Lemma phi_s} is satisfied (see Section \ref{Section: Proof of theorem observer} for details).
Furthermore, we pick $\lambda_f^* = 0.456$, then $T^* = T(L_f,\gamma_f, \lambda_f^*) = 0.45 \ \nicefrac{s}{\epsilon}$. 

For the given $P_f$ and $W_s$, the minimum possible values for $\lambda_1$, $\lambda_2$, $\lambda_3$, $\lambda_4$ and $\lambda_5$ in \eqref{eqn: Vf at slow jumps - Observer} are given by $0.0025$, $0.11$, $0.01$, $0.23$ and $0.01$, respectively. Similarly, for the specified $W_s$, $P_s$, $W_f$ and $P_f$, it can be shown that $b_1 = 14$, $b_2= 0.21$, $b_3 = 2.8$, $a_{\Delta_1} = 0.036$, $a_{\Delta_2} = 0$ and $a_{\Delta_3} = 0.056$.
By selecting $\tau_{\text{miati}}^{s} = 0.149 \ ms$, $\mu = 0.6 a_s \underline{a}_{U_s}$ and $\mu_1 = 0.4 a_s \underline{a}_{U_s}$, we obtain $\epsilon^* = 0.016$, thus the maximum $\tau_{\text{mati}}^f$ is 7 ms. Moreover, we have $\gamma_{v_1} = 8.4$, $\gamma_{v_2} = 8.0$ and $\gamma_{\dot u_s} = 10.0$. In this example, we aim for the maximum MATI estimates, as a result, the gain ultimate bound is large, as we discussed in Remark \ref{Remark: theorem - Observer}.

\begin{figure}[H]
    \centering
    \includegraphics[width = 0.95\linewidth]{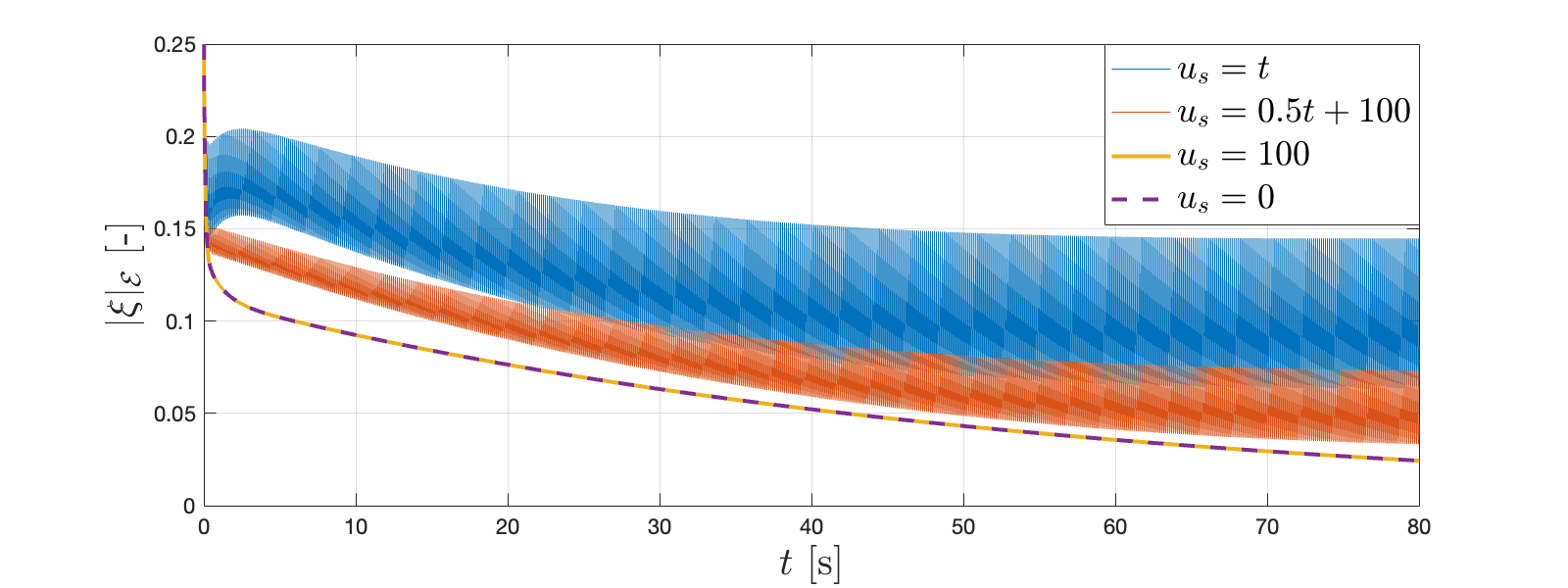}
    \caption{Evolution of $|\xi|_\mathcal{E}$ under different $u_s$}
    \label{fig: Simulation}
\end{figure}

\begin{figure}[H]
    \centering
    \includegraphics[width = 0.95\linewidth]{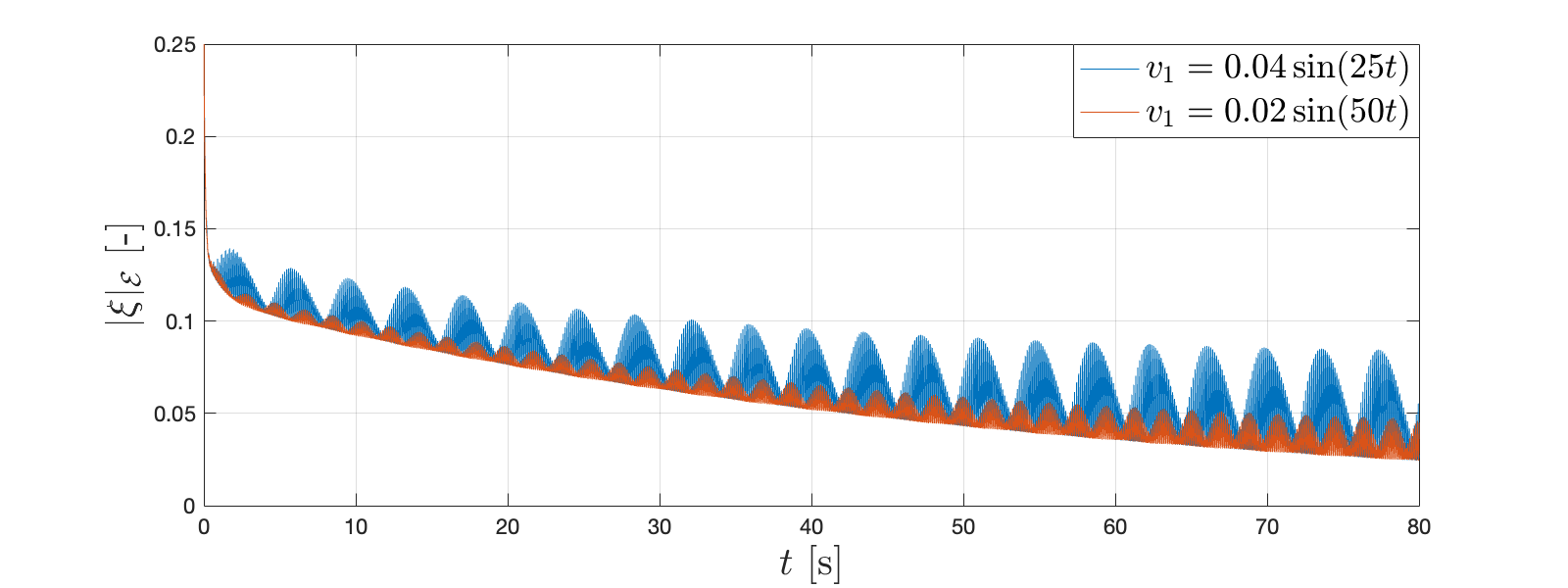}
    \caption{Evolution of $|\xi|_\mathcal{E}$ under different $v_1$}
    \label{fig: Simulation measurement noises}
\end{figure}

To verify the bound \eqref{eqn: DISS}, we conducted simulations for the example presented above using $\epsilon = 0.016$, $\tau_{\text{mati}}^s = 0.15~\text{s}$, and $\tau_{\text{mati}}^f = 7~\text{ms} \leq \epsilon^* T^*$, with initial conditions $x_p(0) = (0.1,0.1)$, $z_p(0) = (-0.2,-0.2)$, and all other states initialized at the origin. The results are shown in Figures~\ref{fig: Simulation} and~\ref{fig: Simulation measurement noises}, where two scenarios are considered: (i) a nonzero control input with no measurement noise, and (ii) zero control input subject to measurement noise.

The first scenario, shown in Figure~\ref{fig: Simulation}, corresponds to $v_1 = v_2 = 0$, where four different control inputs, namely $u_s = t$, $u_s = 0.5t + 100$, $u_s = 100$ and $u_s = 0$, are applied. We observe that the input with the larger time derivative (i.e., $u_s = t$) results in a larger ultimate bound, even though the input with the smaller derivative (i.e., $u_s = 0.5t + 100$) has a greater magnitude. Furthermore, when $\dot u_s = 0$ (as in the cases $u_s = 100$ and $u_s = 0$), $u_s$ has no influence on the observer performance. This behaviour is consistent with the bound in \eqref{eqn: DISS}.
The second scenario, shown in Figure~\ref{fig: Simulation measurement noises}, sets $u_s = 0$ and introduces measurement noise. Two noise signals, $v_1 = 0.04\sin(25t)$ and $v_1 = 0.02\sin(50t)$, are simulated while $v_2 = 0$. The results show that the measurement noise with larger amplitude yields a larger ultimate bound, even though the time derivatives of the two noise signals have equal magnitude. Overall, the system’s response to both control inputs and measurement noise is consistent with the bound in~\eqref{eqn: DISS}, as illustrated in Figures~\ref{fig: Simulation} and~\ref{fig: Simulation measurement noises}.

\section{Conclusion}
This paper investigates the observer design for two-time-scale singularly perturbed networked control systems with linear plants that operate over a single communication channel, subject to both plant inputs and measurement noise. Our analysis demonstrates that under appropriate conditions, the resulting estimation error converges exponentially to a neighborhood of the origin, where the ultimate bound depends explicitly on the magnitudes of the measurement noise and the time derivative of the control input.

\bibliography{ifacconf}             % bib file to produce the bibliography
                                                     % with bibtex (preferred)

%%%%%%%%%%%%%%%%%%%%%%%%%%%%%%%%%%%%%%%%%%%%%%%%%%%%
\appendix
\section{Proof of Theorem \ref{Theorem H - Observer}}    % Each appendix must have a short title.
\label{Section: Proof of theorem observer}

We first analysis the behaviour of systems \eqref{eqn: H_bl - Observer} and \eqref{eqn: H_r - Observer}, then we analysis the trajectory of the overall SPNCS.

\noindent\textbf{Reduced system:}

We denote $W_\star(\kappa_\star, e_\star)$ by $W_\star$ for $\star \in \{s,f \}$. 
For the reduced system \eqref{eqn: H_r - Observer}, we define Lyapunov function $V_s(\delta_x) = \delta_x^\top P_s \delta_x$, where $P_s$ comes from Assumption \ref{Assumption boundary-layer system - Observer}. Then we have $\underline{a}_{Vs} |\delta_x|^2  \leq V_s(\delta_x)  \leq  \overline{a}_{Vs} |\delta_x|^2$, where $\underline{a}_{Vs} = \lambda_\text{min}(P_s)$ and $\overline{a}_{Vs} = \lambda_\text{max}(P_s)$. We can show
\begin{multline}
    \left<  \tfrac{\partial V_s}{\partial \delta_x} , f_{\delta_x}(\delta_x, e_{ys},e_{us},0,0,\hat{v}) \right> 
    =  \delta_x^\top  (P_s A_{11}^s + A_{11}^{s\top}) \delta_x 
    \\
        + \delta_x^\top P_s A_{12}^s e_s + e_s^\top A_{12}^{s\top} P_s \delta_x + 2 \delta_x^\top P_s A_{13}^s \hat{v},
        \label{eqn: V_s dot step 1}
\end{multline}
where
\begin{equation*}
    2 \delta_x^\top P_s A_{13}^s \hat{v} = 2 \delta_x^\top P_s  \begin{bmatrix}A_{14}^r & A_{15}^r \end{bmatrix} \begin{bmatrix}
        \hat{v}_1 \\ \hat v_2.
    \end{bmatrix} 
\end{equation*}
By completion of squares, for any $\eta_{1,1}, \eta_{1,2} >0$, we have
\begin{equation}
    \begin{aligned}
        2 \delta_x^\top P_s A_{14}^r \hat{v}_1 \leq \eta_{1,1} \delta_x^\top P_s^\top P_s \delta_x + a_{V_s}^{v_1} |\hat{v}_1|^2, \\
        2 \delta_x^\top P_s A_{15}^r \hat{v}_2 \leq \eta_{1,2} \delta_x^\top P_s^\top P_s \delta_x + a_{V_s}^{v_2} |\hat{v}_2|^2,
    \end{aligned} 
    \label{eqn: completion of squares 1}
\end{equation}
where $a_{V_s}^{v_1} \coloneqq  \tfrac{1}{\eta_{1,1}} \lambda_{\text{max}}(A_{14}^{r\top} A_{14}^r)$ and $a_{V_s}^{v_2} \coloneqq \tfrac{1}{\eta_{1,2}} \lambda_{\text{max}}(A_{15}^{r\top} \allowbreak A_{15}^r)$. Let $\eta_{1,1} + \eta_{1,2} = \eta_1$, where $\eta_1$ comes from \eqref{eqn: LMI reduced}. By \eqref{eqn: V_s dot step 1} and \eqref{eqn: completion of squares 1}, we can write
\begin{align*}
    &\left<  \tfrac{\partial V_s}{\partial \delta_x} , f_{\delta_x}(\delta_x, e_{ys},e_{us},0,0,\hat{v}) \right> 
    \\
    = & \delta_x^\top  (P_s A_{11}^s + A_{11}^{s\top} + \eta_1 P_s^\top P_s ) \delta_x + \delta_x^\top P_s A_{12}^s e_s
    \\
    & + e_s^\top A_{12}^{s\top} P_s \delta_x 
    + \sigma_{V_s}(|\hat v_1|,|\hat v_2|),
\end{align*}
where $\sigma_{V_s}(|\hat v_1|,|\hat v_2|) = a_{V_s}^{v_1} |\hat v_1|^2 + a_{V_s}^{v_2} |\hat v_2|^2$.
By \eqref{eqn: LMI reduced}, the following inequality hold:
\begin{equation}
    \begin{aligned}
        &\left<  \tfrac{\partial V_s}{\partial \delta_x} , f_{\delta_x}(\delta_x, e_{ys},e_{us},0,0,\hat{v}) \right> 
        \\
    \leq & -a_{\rho_s} |\delta_x|^2 -a_{\rho_s} W_s^2
        - |A_{H_s} \delta_x|^2 
    + \gamma_s^2 W_s^2 + \sigma_{V_s}(|\hat v_1|,|\hat v_2|).
    \end{aligned}
    \label{eqn: reduced system Vs dot}
\end{equation}
Inequality \eqref{eqn: reduced system Vs dot} demonstrates that Assumption \ref{Assumption reduced model - Observer} establishes an $\mathcal{L}_2$ stability of $\mathcal{H}_r$ with gain $\gamma_s$ from $W_s$ to $|A_{H_s}\delta_x|$ \cite{carnevale_stability}.

The function $T$ in \eqref{eqn: function T - Observer} is designed such at for a function $\phi$ whose derivative given by $\dot \phi = -2L\phi - \gamma(\phi^2+1)$ \cite{carnevale_stability}, and the time it takes to decrease from $1/\lambda$ to $\lambda$ is $T(L,\gamma,\lambda)$. 

\begin{lem}
    Let $\tau_{\text{mati}}^s < T(L_s,\gamma_s,\lambda_s)$ be given. There exist $\lambda_s^*\in(\lambda_s, 1)$ and $\eta_s>0$ such that the solution $\phi_s$ to 
    \begin{equation}
        \dot{\phi}_s = -2 L_s \phi_s - \gamma_s[(1+\eta_s)\phi_s^2 + 1],
            \label{eqn: phi_s}
    \end{equation}
    with $\phi_s(0) = 1/\lambda_s^*$ satisfies $\phi_s(\tau_{\text{mati}}^s) = \lambda_s^*$ and $\phi_s(s)\in [\lambda_s^*, 1/\lambda_s^*]$ for all $s\in [0, \tau_{\text{mati}}^s]$.
    \label{Lemma phi_s}
\end{lem}
We define a Lyapunov function $$U_s(\xi_s) = V_s(\delta_x) + \gamma_s \phi_s(\tau_s) W_s^2(\kappa_s, e_s),$$ which was introduced in \cite{carnevale_stability}, where $\phi_s$ and $\lambda_s^*$ come from Lemma \ref{Lemma phi_s}, and $\phi_s(0) = 1/\lambda_s^*$. 
Then by Lemma \ref{Lemma phi_s} and \eqref{eqn: W sandwich bound}, there exist $\underline{a}_{U_s} \coloneqq \min\{\underline{a}_{V_s}, \gamma_s \lambda_s^* \underline{a}_{W_s}^2 \}$ and $\overline{a}_{U_s} \coloneqq \max\{\overline{a}_{V_s}, \gamma_s \tfrac{1}{\lambda_s^*} \overline{a}_{W_s}^2 \}$, such that $\underline{a}_{U_s}|(\delta_x,e_s)|^2 \leq U_s(\xi_s) \leq \overline{a}_{U_s}|(\delta_x,e_s)|^2$.
By \eqref{eqn: reduced system Ws dot} and \eqref{eqn: reduced system Vs dot}, we have
% $
% \big< \tfrac{\partial U_s}{\partial \xi_s}, F_{\xi_s}(\xi_s,0,0,\hat{v},\dot{u}_s) \big>
% = \big<  \tfrac{\partial V_s}{\partial \delta_x} , f_{\delta_x}(\delta_x, e_{ys},e_{us},0,0,\hat{v}) \big> + \tfrac{\partial \phi_s}{\partial \tau_s}\gamma_s W_s^2
% + 2\gamma_s \phi_s(\tau_s) W_s \tfrac{\partial W_s}{\partial e_s}f_{es}(\delta_x,e_{ys}, e_{us},0, 0,\hat{v},\dot u_s)
%         \leq  -a_{\rho_s} (|\delta_x|^2 +W_s^2(\kappa_s, e_s)) - |A_{H_s} \delta_x|^2 + \gamma_s^2 W_s^2
%             + a_{v_s} |\hat{v}|^2 + \tfrac{\partial \phi_s}{\partial \tau_s}\gamma_s W_s^2 
%             + 2\gamma_s \phi_s(\tau_s) W_s 
%             [L_s W_s + |A_{H_s} \delta_x| + M_s \sigma_{w_s}(|\hat{v}|, |\dot u_s|)]
% $.
\begin{equation*}
    \begin{aligned}
        &\left< \tfrac{\partial U_s}{\partial \xi_s}, F_{\xi_s}(\xi_s,0,0,\hat{v},\dot{u}_s) \right>
        \\
        =& \left<  \tfrac{\partial V_s}{\partial \delta_x} , f_{\delta_x}(\delta_x, e_{ys},e_{us},0,0,\hat{v}) \right> 
            \\
            &+ \tfrac{\partial \phi_s}{\partial \tau_s}\gamma_s W_s^2
        + 2\gamma_s \phi_s W_s \tfrac{\partial W_s}{\partial e_s}f_{es}(\delta_x,e_{ys}, e_{us},0, 0,\hat{v},\dot u_s)
        \\
        \leq & -a_{\rho_s} (|\delta_x|^2 +W_s^2(\kappa_s, e_s)) - |A_{H_s} \delta_x|^2 + \gamma_s^2 W_s^2 
            \\
            &+ \sigma_{V_s}(|\hat v_1|,|\hat v_2|) + \tfrac{\partial \phi_s}{\partial \tau_s}\gamma_s W_s^2 
            \\
            &+ 2\gamma_s \phi_s W_s 
            [L_s W_s + |A_{H_s} \delta_x| + M_s \sigma_{w_s}(|\hat{v}_1|,|\hat{v}_2|, |\dot u_s|)].
    \end{aligned}
\end{equation*} 
By the definition of $\sigma_{w_s}$ and completion of squares, for any $\eta_{s1},\eta_{s2}, \eta_{s3}>0$, we have 
\begin{multline*}
    2\gamma_s \phi_s W_s M_s\sigma_{w_s} \leq (\eta_{s1}+\eta_{s2}+ \eta_{s3}) \gamma_s^2 \phi_s^2 W_s^2 
    \\
    + \left(\tfrac{{a_{w_s}^{v_1}}^2 }{\eta_{s1}}|\hat{v}_1|^2  +\tfrac{{a_{w_s}^{v_2}}^2} {\eta_{s2}}|\hat{v}_2|^2 +  \tfrac{M_s^2}{\eta_{s2}}|\dot u_s|^2 \right).
\end{multline*}
Let $\eta_{s1}+\eta_{s2}+ \eta_{s3} = \eta_s$, where $\eta_s$ comes from Lemma \ref{Lemma phi_s}. Moreover, we have $$2\gamma_s \phi_s W_s |A_{H_s} \delta_x| \leq  \gamma_s^2 \phi_s^2 W_s^2 + |A_{H_s} \delta_x|^2.$$ Hence, by definition of $\dot \phi_s$ in \eqref{eqn: phi_s}, we have that for reduced system,
\begin{equation}
    \big<\! \tfrac{\partial U_s}{\partial \xi_s},F_{\xi_s}(\xi_s, 0, 0, \hat{v}, \dot{u}_s) \! \big>\! \leq \! -a_s |(\delta_x, e_s)|^2 + \sigma_s(|\hat{v}_1|,|\hat{v}_2|, |\dot u_s|), 
    \label{eqn: U_s dot}
\end{equation}
where $a_s \coloneqq a_{\rho_s} \text{min}\{1, \underline{a}_{W_s}^2 \}$ and 
\begin{multline*}
    \sigma_s(|\hat{v}_1|,|\hat{v}_2|, |\dot u_s|) \coloneqq   \left(\tfrac{{a_{w_s}^{v_1}}^2 }{\eta_{s1}}|\hat{v}_1|^2 +\tfrac{{a_{w_s}^{v_2}}^2} {\eta_{s2}}|\hat{v}_2|^2 +  \tfrac{M_s^2}{\eta_{s2}}|\dot u_s|^2 \right)  
    \\
    + \sigma_{V_s}(|\hat v_1|,|\hat v_2|).
\end{multline*}
By Lemma \ref{Lemma phi_s}, at transmission times of the reduced system (i.e., $t \in \mathcal{T}^s$), we have 
\begin{equation*}
    \begin{aligned}
        U_s(\xi_s^+) =&  U_s\big((\delta_x, h_s(\kappa_s, e_s), 0, \kappa_s + 1)\big) 
        \\
        =& V_s(\delta_x) + \gamma_s \phi_s(0) W_s^2(\kappa_s + 1, h_s(\kappa_s, e_s))
        \\
        \leq & V_s(\delta_x) + \tfrac{\gamma_s}{\lambda_s^*} \lambda_s W_s^2(\kappa_s, e_s) 
        \\
        \leq & U_s(\xi_s) .
    \end{aligned}
\end{equation*}

\noindent \textbf{Boundary-layer system:}

For boundary-layer system \eqref{eqn: H_bl - Observer}, let $V_f(\delta_y) = \delta_y^\top P_f \delta_y$, where $P_f$ comes from Assumption \ref{Assumption boundary-layer system - Observer}, then we have $\underline{a}_{Vf} |\delta_y|^2  \leq V_f(\delta_y)  \leq  \overline{a}_{Vf} |\delta_y|^2$, where $\underline{a}_{Vf} = \lambda_\text{min}(P_f)$ and $\overline{a}_{Vf} = \lambda_\text{max}(P_f)$. Similar to  \eqref{eqn: reduced system Vs dot}, we can show
%$\big< \tfrac{\partial V_f}{\partial \delta_y} , g_{\delta_y}(\delta_x, e_{ys},e_{us},\delta_y,e_f,\hat{v}, \dot v_1, 0)  \big> \leq -a_{\rho_f} |\delta_y|^2 - a_{\rho_f} W_f^2(\kappa_f, e_f) - |A_{H_f} \delta_y|^2 + \gamma_f^2 W_f^2(\kappa_f, e_f).$
\begin{multline*}
    \left< \tfrac{\partial V_f}{\partial \delta_y} , g_{\delta_y}(\delta_x, e_{ys},e_{us},\delta_y,e_f,\hat{v}, \dot v_1, 0)  \right> \leq 
    \\
    -a_{\rho_f} |\delta_y|^2  - a_{\rho_f} W_f^2(\kappa_f, e_f) 
    - |A_{H_f} \delta_y|^2 + \gamma_f^2 W_f^2(\kappa_f, e_f).
\end{multline*}
We define Lyapunov function 
$$U_f(\xi_f) = V_f(\delta_y) + \gamma_f \phi_f(\tau_f) W_f^2(\kappa_f, e_f),$$
where $\tfrac{\partial {\phi}_f}{\partial \sigma} = -2 L_f \phi_f - \gamma_f(\phi_f^2 + 1)$, $\phi_f(0) = 1/\lambda_f^*$ and $\lambda_f^*$ comes from Theorem \ref{Theorem H - Observer}.
Given $T^* \leq T(L_f,\gamma_f,\lambda_f^*)$, we have $\phi_f(s)\in [\lambda_f^*, 1/\lambda_f^*]$ for all $s \in [0,T^*]$. Then by \eqref{eqn: W sandwich bound}, we can find $\underline{a}_{U_f} \coloneqq \min\{\underline{a}_{V_f}, \gamma_f \lambda_f^* \underline{a}_{W_f}^2 \}$ and $\overline{a}_{U_f} \coloneqq \max\{\overline{a}_{V_f}, \gamma_f \tfrac{1}{\lambda_f^*} \overline{a}_{W_f}^2 \}$ such that $\underline{a}_{U_f}|(\delta_y,e_f)|^2 \leq U_f(\xi_f) \leq \overline{a}_{U_f}|(\delta_y,e_f)|^2$.
Along the same lines as for the reduced system, we can show %$\big< \tfrac{\partial U_f}{\partial \xi_f}, F_{\xi_f}(\xi_s,0,0,\hat{v},\dot{u}_s) \big> \leq -a_f |(\delta_y, e_f)|^2$,
\begin{equation*}
    \big< \tfrac{\partial U_f}{\partial \xi_f}, F_{\xi_f}(\xi_s,0,0,\hat{v},\dot{u}_s) \big> \leq -a_f |(\delta_y, e_f)|^2,
    \label{eqn: U_f dot - Observer} 
\end{equation*}
where $a_f \coloneqq a_{\rho_f} \text{min}\{1, \underline{a}_{W_f}^2 \}$. Moreover, at transmission time of $\mathcal{H}_{bl}$ (i.e., $t \in \mathcal{T}^f$), we can show $U_f(\xi_f^+) \leq U_f(\xi_f$.

\noindent \textbf{Overall SPNCS:}

Now we focus on the full system $\mathcal{H}$. Remind that at transmission time of slow signals, we have $\delta_y$ updating according to \eqref{eqn: Jump of y at slow transmission - Observer}. Then by \eqref{eqn: W sandwich bound} and definition of $V_f$, we can find $\lambda_1, \lambda_2, \lambda_3, \lambda_4, \lambda_5 \geq 0$ such that 
\begin{equation}
\begin{aligned}
    &V_f(h_y(\kappa_s, e_{ys}, e_{us}, \tilde e_{p_s}, \delta_y, \hat v_1, v_1)) - V_f(\delta_y) 
    \\
    \leq & \lambda_1 W_s^2(\kappa_s, e_s) + \lambda_2 \sqrt{W_s^2(\kappa_s, e_s)V_f(\delta_y)}  
    \\
    & + \lambda_3 |\hat{v}|^2 + \lambda_4 \sqrt{V_f}|\hat v| + \lambda_5 W_s |\hat v|.
    \end{aligned}
    \label{eqn: Vf at slow jumps - Observer}
\end{equation}

Let $a_{\psi_s} \coloneqq \sqrt{\underline{a}_{U_s}}$, $a_{\psi_f} \coloneqq \sqrt{\underline{a}_{U_f}}$ and 
\begin{equation}
    \mu \in (0, a_s a_{\psi_s}^2), \quad \mu_1 \in (0, \mu). \label{eqn: mu and mu1 - Observer}
\end{equation}
Let
\begin{equation}
    \tilde \lambda \in ( e^{-\mu_1 \tau_{\text{miati}}^s}, 1), \label{eqn: tilde lambda - Observer}
\end{equation}
where $\tau_{\text{miati}}^{s}$ comes from Theorem \ref{Theorem H - Observer}. 
We define %$ d \coloneqq \big(\tfrac{-b+\sqrt{b^2-4ac}}{2a}\big)^2$,
\begin{equation}
    d \coloneqq \left(\tfrac{-b+\sqrt{b^2-4ac}}{2a}\right)^2,
    \label{eqn: d - Observer}
\end{equation}
where $a = \max \{ \tfrac{\lambda_1+\nicefrac{\lambda_5}{2}}{\gamma_s \lambda_s^*}, \tfrac{\lambda_4}{2} \}$, $b= \tfrac{\lambda_2}{2} \max \{ \tfrac{\lambda_1}{\gamma_s \lambda_s^*}, 1 \}$ and $c = 1 - \tilde \lambda e^{\mu_1 \tau_{\text{miati}}^s}$. Let 
\begin{equation}
    a_d \coloneqq 1 + a\cdot d + b \sqrt{d}.
    \label{eqn: a_d - Observer}
\end{equation}
We define a composite Lyapunov function 
$$U(\xi_s,\xi_f)\coloneqq U_s(\xi_s) + d U_f(\xi_f),$$ 
and we can find $\underline{a}_{U}, \overline{a}_U > 0$ such that $\underline{a}_{U}|\xi|_{\mathcal{E}}^2 \leq U(\xi) \leq \overline{a}_U |\xi|_{\mathcal{E}}^2$. We can show at a fast transmission, $U(\xi_s^+, \xi_f^+) \leq U(\xi_s, \xi_f)$. During flow, the time derivative of $U$ along the trajectories of $\mathcal{H}$ is given by
\begin{equation*}
    \begin{aligned}
        &\tfrac{\partial U(\xi_s,\xi_f)}{\partial t}  
        \\
        =& \tfrac{\partial U_s}{\partial \xi_s} F_{\xi_s}(\xi_s, \delta_y, e_f, \hat{v},  \dot{u}_s) 
            + \tfrac{d}{\epsilon}  \tfrac{\partial U_f}{\partial \xi_f} F_{\xi_f}(\xi_s,\xi_f,\hat{v}, \dot{u}_s, \epsilon) 
        \\
        =&  \tfrac{\partial U_s}{\partial \xi_s} F_{\xi_s}(\xi_s, 0, 0, \hat{v}, \dot{u}_s) + \tfrac{d}{\epsilon} \tfrac{\partial U_f}{\partial \xi_f} F_{\xi_f}(\xi_s, \xi_f, \hat{v}, \dot{u}_s,0) \\
            &+\tfrac{\partial U_s}{\partial \xi_s} \left[F_{\xi_s}(\xi_s, \delta_y, e_f, \hat{v}, \dot{u}_s) -  F_{\xi_s}(\xi_s, 0, 0, \hat{v}, \dot{u}_s)\right]
            \\
            &+ \tfrac{d}{\epsilon}  \tfrac{\partial U_f}{\partial \xi_f} \left[F_{\xi_f}(\xi_s,\xi_f,\hat{v}, \dot{u}_s, \epsilon) - F_{\xi_f}(\xi_s, \xi_f, \hat{v}, \dot{u}_s,0) \right].
    \end{aligned}
\end{equation*}
%
% \begin{equation*}
%     \begin{aligned}
%         &\tfrac{\partial U(\xi_s,\xi_f)}{\partial t}  
%         \\
%         =& \tfrac{\partial U_s}{\partial \xi_s} F_{\xi_s}(\xi_s, \delta_y, e_f, \hat{v},  \dot{u}_s) 
%             + \tfrac{d}{\epsilon}  \tfrac{\partial U_f}{\partial \xi_f} F_{\xi_f}(\xi_s,\xi_f,\hat{v}, \dot{u}_s, \epsilon) 
%     \end{aligned}
% \end{equation*}
% %
% \begin{equation*}
%     \begin{aligned}
%         =&  \tfrac{\partial U_s}{\partial \xi_s} F_{\xi_s}(\xi_s, 0, 0, \hat{v}, \dot{u}_s) + \tfrac{d}{\epsilon} \tfrac{\partial U_f}{\partial \xi_f} F_{\xi_f}(\xi_s, \xi_f, \hat{v}, \dot{u}_s,0) \\
%             &+\tfrac{\partial U_s}{\partial \xi_s} \left[F_{\xi_s}(\xi_s, \delta_y, e_f, \hat{v}, \dot{u}_s) -  F_{\xi_s}(\xi_s, 0, 0, \hat{v}, \dot{u}_s)\right]
%             \\
%             &+ \tfrac{d}{\epsilon}  \tfrac{\partial U_f}{\partial \xi_f} \left[F_{\xi_f}(\xi_s,\xi_f,\hat{v}, \dot{u}_s, \epsilon) - F_{\xi_f}(\xi_s, \xi_f, \hat{v}, \dot{u}_s,0) \right]
%     \end{aligned}
% \end{equation*}

By definition of $U_s$, $F_{\xi_s}$ and Assumption \ref{Assumption UGES protocol} (i.e., $\left| \tfrac{\partial W_s}{\partial e_s} \right| \leq M_s $), there exists $b_1 \geq 0$ such that 
\begin{multline*}
    \tfrac{\partial U_s}{\partial \xi_s} [F_{\xi_s}(\xi_s, \delta_y, e_f, \hat{v}, \dot{u}_s) -  F_{\xi_s}(\xi_s, 0, 0, \hat{v}, \dot{u}_s)] \leq 
    \\
    b_1 |(\delta_x, e_s)|\cdot  |(\delta_y, e_f)|.
\end{multline*}
Similarly, by definition of $U_f$, $F_{\xi_f}$, Assumption \ref{Assumption UGES protocol} and completion of squares, we can show there exist $b_2,b_3,a_{\Delta_1},a_{\Delta_2},\allowbreak a_{\Delta_3}\geq0$ such that 
\begin{multline*}
    \tfrac{1}{\epsilon}  \tfrac{\partial U_f}{\partial \xi_f} \left[F_{\xi_f}(\xi_s,\xi_f,\hat{v}, \dot{u}_s, \epsilon) - F_{\xi_f}(\xi_s, \xi_f, \hat{v}, \dot{u}_s,0) \right] 
    \\
    \leq b_2 |(\delta_x, e_s)|\cdot|(\delta_y, e_f)|+b_3|(\delta_y, e_f)|^2 + \Delta(|\hat{v}_1|,|\hat{v}_2|,|\dot u_s|),
\end{multline*}
%$\tfrac{1}{\epsilon}  \tfrac{\partial U_f}{\partial \xi_f} \left[F_{\xi_f}(\xi_s,\xi_f,\hat{v}, \dot{u}_s, \epsilon) - F_{\xi_f}(\xi_s, \xi_f, \hat{v}, \dot{u}_s,0) \right] \leq b_2 |(\delta_x, e_s)|\cdot|(\delta_y, e_f)|+b_3|(\delta_y, e_f)|^2 + \Delta(|\hat{v}|,|\dot u_s|)$, 
where $\Delta(|\hat{v}_1|,|\hat{v}_2|,|\dot u_s|) \coloneqq a_{\Delta_1} |\hat{v}_1|^2 + a_{\Delta_2} |\hat{v}_2|^2 + a_{\Delta_3} |\dot u_s|^2$. 
Then we have 
$$\tfrac{\partial U(\xi_s,\xi_f)}{\partial t} \leq -\left[\begin{smallmatrix}\sqrt{U_s(\xi_s)} \\ \sqrt{U_f(\xi_f)} \end{smallmatrix}\right]^\top \Lambda \left[\begin{smallmatrix}\sqrt{U_s(\xi_s)} \\ \sqrt{U_f(\xi_f)} \end{smallmatrix}\right] 
+\Gamma(|\hat{v}_1|, |\hat{v}_2|,|\dot u_s|),$$
where 
\begin{equation*}
    \Lambda \coloneqq \left[\begin{smallmatrix}
            a_s a_{\psi_s}^2 & - \tfrac{1}{2}(b_1 + d b_2) a_{\psi_s}a_{\phi_f} \\ - \tfrac{1}{2}(b_1 + d b_2)a_{\psi_s}a_{\psi_f} & d(\frac{a_f}{\epsilon}-b_3) a_{\psi_f}^2
        \end{smallmatrix} \right],
\end{equation*}
and 
\begin{equation*}
\Gamma(|\hat{v}_1|,|\hat{v}_2|, |\dot u_s|) \coloneqq \sigma_s(|\hat{v}_1|,|\hat{v}_2|, |\dot u_s|) + d\cdot\Delta(|\hat{v}_1|,|\hat{v}_2|, |\dot u_s|).
\end{equation*}
We define
% $$
%     \epsilon^* = \big(\tfrac{a_{\psi_f}}{a_f d}(\tfrac{(b_1 + db_2)^2 a_{\psi_s}^2 a_{\psi_f}^2}{4(a_s a_{\psi_s}^2 - \mu)}+\mu d) + \tfrac{b_3 a_{\psi_f}^2}{a_f} \big)^{-1},
% $$
$$    \epsilon^* = \left(\tfrac{1}{d a_f  a_{\psi_f}^2}\Big(\tfrac{(b_1 + db_2)^2 a_{\psi_s}^2 a_{\psi_f}^2}{4(a_s a_{\psi_s}^2 - \mu)}+\mu d\Big) + \tfrac{b_3}{a_f} \right)^{-1},$$
then, by definition of $d$ in \eqref{eqn: d - Observer} and $\mu$ in \eqref{eqn: mu and mu1 - Observer}, we have $\Lambda \geq \mu \left[ \begin{smallmatrix}
   1 & 0 \\ 0 & d 
\end{smallmatrix}\right]$, and by definition of $\mu_1$ in \eqref{eqn: mu and mu1 - Observer}, we have 
\begin{equation*}
    \tfrac{\partial U(\xi_s,\xi_f)}{\partial t} \leq - \mu_1 U(\xi_s, \xi_f), \quad \forall U(\xi_s,\xi_f)\geq \tfrac{\Gamma(|\hat{v}_1|,|\hat{v}_2|, |\dot u_s|)}{\mu-\mu_1}.
\end{equation*}
%$\tfrac{\partial U(\xi_s,\xi_f)}{\partial t} \leq - \mu_1 U(\xi_s, \xi_f)$ for all $U(\xi_s,\xi_f)\geq \tfrac{\Gamma(|\hat{v}|, |\dot u_s|)}{\mu-\mu_1}$. 
Let $t_{j_k^s} \in \mathcal{T}^s$ correspond to the transmission instance of the ${k^s}^\text{th}$ slow transmission, $j_k^s$ denotes the number of transmissions including the ${k^s}^\text{th}$ slow transmission at time $t_{j_k^s}$, and we denote $U(\xi_s(t,j), \xi_f(t,j))$ by $U(t,j)$. Then, by comparison principle and the fact that $U$ is non-increasing at fast transmission, we have that for all $t_{j_k^s},t_{j_{k+1}^s} \in \mathcal{T}^s$, if $U(t_{j_k^s}, j_k^s) \geq \tfrac{\Gamma(|\hat{v}_1|,|\hat{v}_2|, |\dot u_s)}{\mu-\mu_1}$, then 
%$U(s,i) \leq U(t_{j_k^s}, j_k^s) \exp  \big(-\mu_1(s-t_{j_k^s}) \big)$
\begin{equation*}
    U(s,i) \leq U(t_{j_k^s}, j_k^s) \exp  \big(-\mu_1(s-t_{j_k^s}) \big) \label{eqn: Exponential U flow - Observer}, 
\end{equation*}
whenever $U(s,i) \geq \tfrac{\Gamma(|\hat{v}_1|,|\hat{v}_2|, |\dot u_s)}{\mu-\mu_1}$, as well as $(t_{j_k^s}, j_k^s) \preceq  (s,i)  \preceq (t_{j_{k+1}^s}, j_{k+1}^s - 1)$.

At slow transmissions, we have 
\begin{equation*}
    U(\xi_s^+,\xi_f^+) \leq U_s(\xi_s) + d V_f(h_y(\kappa_s, e_{ys}, e_{us}, \tilde e_{p_s}, \delta_y, \hat v_1, v_1)).
\end{equation*}
By adding and subtracting $V_f(\xi_f)$, we have 
\begin{multline*}
    U(\xi_s^+, \xi_f^+) \leq U(\xi_s,\xi_f) 
    + d \Big(\lambda_1 W_s^2(\kappa_s, e_s) 
    \\
    + \lambda_2 \sqrt{W_s^2(\kappa_s, e_s)V_f(\delta_y)} 
    + \lambda_3 |\hat{v}_1|^2 + \lambda_4 \sqrt{V_f}|\hat v_1| + \lambda_5 W_s |\hat v_1| \Big),
\end{multline*}
%$U(\xi_s^+, \xi_f^+) \leq U(\xi_s,\xi_f) + d \big(\lambda_1 W_s^2(\kappa_s, e_s) + \lambda_2 \sqrt{W_s^2(\kappa_s, e_s)V_f(\delta_y)} + \lambda_3 |\hat{v}|^2 + \lambda_4 \sqrt{V_f}|\hat v| + \lambda_5 W_s |\hat v| \big)$, 
and by completion of square, we can obtain 
\begin{equation*}
    U(\xi_s^+, \xi_f^+) \leq a_d U(\xi_s,\xi_f) + a_v |\hat v_1|^2,
\end{equation*}
where $a_d$ is defined in \eqref{eqn: a_d - Observer} and
$$a_v  \coloneqq d\lambda_3 + \tfrac{\lambda_4+d\lambda_5}{2}.$$
Then if
$U(s,i)\geq \tfrac{\Gamma(|\hat v_1|,|\hat v_2|, |\dot u_s|)}{\mu-\mu_1}$ for all $(t_{j_k^s}, j_k^s) \preceq (s,i) \preceq (t_{j_{k+1}^s}, j_{k+1}^s - 1)$, then 
\begin{equation*}
    U(t^s_{k+1}, j^s_{k+1}) \leq \left(a_d U(t_{j_k^s}, j_k^s)+ a_v |\hat v_1|^2 \right)  e^{-\mu_1 \tau_{\text{miati}}^s}.
\end{equation*}
By definition of $\tilde \lambda$ in \eqref{eqn: tilde lambda - Observer}, we have $a_d  e^{-\mu_1 \tau_{\text{miati}}^s} = \tilde \lambda$. Let $\lambda \in (\tilde \lambda, 1)$, then if $U(s,i)\geq \tilde \Gamma(|\hat v_1|,|\hat v_2|, |\dot u_s|)$ for all $(t_{j_k^s}, j_k^s) \preceq (s,i) \preceq (t_{j_{k+1}^s}, j_{k+1}^s-1)$, we have 
$$U(t^s_{k+1}, j^s_{k+1}) \leq \lambda U(t_{j_k^s}, j_k^s),$$
where $\tilde \Gamma(|\hat v_1|,|\hat v_2|, |\dot u_s|) \coloneqq \max \{  \tfrac{a_v |\hat v_1|^2}{\lambda - \tilde \lambda}, \tfrac{\Gamma(|\hat v_1|,|\hat v_2|, |\dot u_s|)}{\mu-\mu_1} \}$.

Following a similar proof as in \cite{Weixuan_Journal} and by comparison principle, we can show that
%$    |\xi(t,j)|_{\mathcal{E}} \leq \big(\tfrac{a_d \overline{a}_U}{\lambda \underline{a}_U} \big)^{\nicefrac{1}{2}}  |\xi(0,0)|_{\mathcal{E}}  \exp\big(-\tfrac{\ln{(\nicefrac{1}{\lambda})}}{2 \tau_{\text{mati}}^s}t\big) + \gamma_{\hat v}\|\hat{v}\|_{(t,j)} + \gamma_{\dot u_s}\|\dot u_s\|_{(t,j)}$,
\begin{multline*}
    |\xi(t,j)|_{\mathcal{E}} \leq \big(\tfrac{a_d \overline{a}_U}{\lambda \underline{a}_U} \big)^{\nicefrac{1}{2}}  |\xi(0,0)|_{\mathcal{E}}  \text{exp}\left(-\tfrac{\ln{(\nicefrac{1}{\lambda})}}{2 \tau_{\text{mati}}^s}t \right)
    \\
    + \gamma_{\hat v}\|\hat{v}\|_{(t,j)} + \gamma_{\dot u_s}\|\dot u_s\|_{(t,j)},
\end{multline*}
where 
\begin{align*}
    \gamma_{\hat v_1} &= \left(\tfrac{a_d}{\underline{a}_U} \max \left\{\tfrac{a_v}{\lambda - \tilde \lambda}, \tfrac{1}{\mu - \mu_1}\big(\tfrac{{a_{ws}^{v_1}}^2}{\eta_{s1}} + a_{V_s}^{v_1} + d a_{\Delta_1} \big) \right\} \right)^{\nicefrac{1}{2}}    
    \\
    \gamma_{\hat v_2} &= \left(\tfrac{a_d}{\underline{a}_U(\mu - \mu_1)}\big(\tfrac{{a_{ws}^{v_2}}^2}{\eta_{s2}} + a_{V_s}^{v_2} + d a_{\Delta_2}\big) \right)^{\nicefrac{1}{2}} 
    \\
    \gamma_{\dot u_s} &= \big(\tfrac{a_d}{\underline{a}_U}(\tfrac{M_s^2}{\eta_{s3}} + d a_{\Delta_3}  ) \big)^{\nicefrac{1}{2}}
\end{align*}
Moreover, we can always find $\gamma_{v_\ell} \geq 0$ such that $\gamma_{\hat v_\ell}\|\hat{v}_\ell\|_{(t,j)} \leq \gamma_{v_\ell}\|{v_\ell}\|_{(t,j)}$ for $\ell \in \{1,2 \}$. For instance, $\gamma_{v_\ell} = \gamma_{\hat v_\ell}$ in sampled-data system. Finally, we prove Theorem \ref{Theorem H - Observer} by using equation (A.25) in \cite{Weixuan_Journal}.

 %%%%%%%%%%%%%%%%%%%%%%%%%%%%%%%%%%%%%%%%%%%%%%%%%%%%                                                                        % in the appendices.
\end{document}